%% file: specpap.tex
\documentclass{aa-package/aa}  
\usepackage{graphicx}
\usepackage{txfonts}
\usepackage{url}

\newcommand{\fn}[1]{\footnote{#1}}
\newcommand{\radec}[9]{\mbox{R.A.$\,$#1$^{\mathrm{h}}$#2$^{\mathrm{m}}$#3\fs#4, Dec.$\,$#5\degr#6\arcmin#7\farcs#8~(#9)}}

\newcommand{\kmps}{\mbox{$\rm km\,s^{-1}$}}
\newcommand{\mps}{\mbox{$\rm m\,s^{-1}$}}
\newcommand{\Tsys}{\mbox{$T\!\!_\mathrm{sys}$}}
\newcommand{\XIIIco}{\mbox{$^{13}$CO}}

\newcommand{\chIIIoh}{\mbox{CH$_{3}$OH}}
\newcommand{\dimetet}{\mbox{(CH$_{3}$)$_{2}$OH}}
\newcommand{\metform}{\mbox{CH$_{3}$OCHO}}
\newcommand{\rem}[1]{\hspace{-0ex}}

\newcommand{\vlsr}{\mbox{$v_{\rm LSR}$}}

\newcommand{\chIIIcn}{\mbox{CH$_{3}$CN}}

\newcommand{\Kkmps}{\mbox{K$\,$\kmps}}

\newcommand{\pscm}{\mbox{cm$^{-2}$}}

\newcommand{\htwoo}{\mbox{H$_{2}$O}}

\newcommand{\sotwo}{\mbox{SO$_{2}$}}

\newcommand{\jtrans}[2]{\mbox{$J\!=\!#1\!-\!#2$}}

\begin{document}
   \title{A spectral line survey of Orion KL in the bands 486--492 and 541--577 GHz with the Odin\thanks{Odin is a Swedish-led satellite project funded jointly by the Swedish National Space Board (SNSB), the Canadian Space Agency (CSA), the National Technology Agency of Finland (Tekes), and the Centre National d'\'Etudes Spatiales (CNES, France). The Swedish Space Corporation (SSC) was the industrial prime contractor and is also responsible for the satellite operation.}\ satellite
}

   \subtitle{I. The observational data}

   \author{A.O.H.~Olofsson\inst{1,2}
          \and
          C.M.~Persson\inst{1} 
	  \and
	  N.~Koning\inst{3}	  
          \and
          P.~Bergman\inst{1,4}
          \and
          P.F.~Bernath\inst{5,6,7}
          \and
	  J.H.~Black\inst{1}
          \and
          U.~Frisk\inst{8}
	  \and
	  \\W.~Geppert\inst{9}
	  \and
          T.I.~Hasegawa\inst{3,10}
          \and
          \AA.~Hjalmarson\inst{1}
          \and
	  S.~Kwok\inst{3,11}
	  \and
	  B.~Larsson\inst{12}
          \and
	  A.~Lecacheux\inst{13}
          \and
	  A.~Nummelin\inst{14}
	  \and
          \\M.~Olberg\inst{1}
          \and
          Aa.~Sandqvist\inst{12}
	  \and
	  E.S.~Wirstr\"om\inst{1}
          }

   \offprints{A.O.H. Olofsson}

   \institute{Onsala Space Observatory (OSO),
              SE-439 92 Onsala, Sweden, \email{henrik@oso.chalmers.se}
         \and
              LERMA, Observatoire de Paris, 61 Av. de l'Observatoire,
              75014 Paris, France
         \and
             Department of Physics and Astronomy, University of Calgary,
	     Calgary, AB T2N 1N4, Canada
	 \and
             European Southern Observatory, Alonso de Cordova 3107,
             Vitacura, Casilla 19001, Santiago, Chile
         \and
             Department of Chemistry, University of Arizona, Tucson, AZ 85721
	 \and	
             Department of Chemistry, University of Waterloo, Waterloo,
             ON N2L 3G1, Canada
         \and
             Department of Chemistry, University of York, Heslington,
             York YO10 5DD, United Kingdom
         \and
	     Swedish Space Corporation, PO Box 4207, 171 04 Solna, Sweden
	 \and
	     Molecular Physics Division, Department of Physics,
             Stockholm University AlbaNova, 10691 Stockholm, Sweden
         \and
             Institute of Astronomy and Astrophysics, Academia Sinica,
             P.O. Box 23-141, Taipei 106, Taiwan, R.O.C.
	 \and
             Department of Physics, University of Hong Kong, Hong Kong, China
         \and
             Stockholm Observatory, AlbaNova University Center,
             106 91 Stockholm, Sweden
	 \and
             LESIA, Observatoire de Paris, Section de Meudon, 5,
             Place Jules Janssen, 92195 MEUDON CEDEX, France
	 \and
             Computer science and engineering,
             Chalmers University of Technology,
             SE-412 96 G\"oteborg, Sweden	
             }

   \date{Received Month MM, 9999; accepted Month MM, 9999}

 
  \abstract
{}
   {Spectral line surveys are useful since they allow
identification of new molecules and new lines in uniformly calibrated
data sets. The subsequent multi-transition analysis will provide improved
knowledge of molecular abundances, cloud temperatures and densities, and
may also reveal previously unsuspected blends of molecular lines, which
otherwise may lead to erroneous conclusions. Nonetheless, large portions of
the sub-millimetre spectral regime remain unexplored due to severe absorptions
by H$_{2}$O and O$_{2}$\ in the terrestrial atmosphere. The purpose of
the measurements presented here is to cover wavelength regions at and around
0.55 mm -- regions largely unobservable from the ground.}  
   {Using the Odin astronomy/aeronomy satellite, we performed the first
spectral survey of the Orion KL molecular cloud core in the bands
486--492 and 541--576 GHz with rather uniform sensitivity
(22--25 mK baseline noise). Odin's 1.1 m size telescope, equipped with four
cryo-cooled tuneable mixers connected to broad band
spectrometers, was used in a satellite position-switching mode. Two
mixers simultaneously observed different 1.1 GHz
bands using frequency steps of 0.5 GHz (25 hours each).
An on-source integration time of 20 hours was achieved for most bands.
The entire campaign consumed $\sim$1100 orbits, each containing one
hour of serviceable astro-observation.}
   {We identified 280 spectral lines from 38 known interstellar
molecules (including isotopologues) having intensities in the range 80
to 0.05 K. An additional 64 weak lines remain unidentified.
Apart from the ground state rotational 1$_{1,0}$--1$_{0,1}$\
transitions of \textit{ortho}-H$_{2}$O, H$_{2}^{18}$O and H$_{2}^{17}$O,
the high energy 6$_{2,4}$--7$_{1,7}$\ line of
\textit{para}-H$_{2}$O ($E_{u}$=867$\,$K)
and the HDO(2$_{0,2}$--1$_{1,1}$) line have been observed, as well as the
1$_{0}$--0$_{1}$\ lines from NH$_{3}$\ and its rare isotopologue
$^{15}$NH$_{3}$.
We suggest assignments for some unidentified features, notably the new
interstellar molecules ND and SH$^{-}$. Severe blends have been detected
in the line wings of the H$_{2}^{18}$O, H$_{2}^{17}$O and $^{13}$CO lines
changing the true linewidths of the outflow emission.}
   {}


   \maketitle
%

\section{Introduction}
\label{introsec}

Being the most popular target for spectral line surveys, in the Orion KL
position OMC-1 has been the focal point of at least $\sim$20 observational
efforts in the mm and submm bands over
the last 20 years, starting with
Johansson et al. (\cite{johansson84}, \cite{johansson85}),
and in the frequency range 72--91 GHz. White et al. (\cite{white03})
provide an extensive list of this earlier work in their introduction.
Using the James Clerk Maxwell Telescope (JCMT), White et al. (\cite{white03})
surveyed the bands 455--469 and
492--507 GHz, surrounding the lowest frequency range of our Odin spectral scan
(486--492 GHz). The frequency range 607--725 GHz, just above the Odin
spectral scan band 542--576 GHz, has been surveyed by Schilke et al.
(\cite{schilke01}) using the Caltech Submillimeter Observatory (CSO).

More recent additions include 159.7--164.7 GHz (Lee \& Cho \cite{lee02}),
795--903 GHz (Comito et al. \cite{comito05}), 260--328 GHz
(Yoshida \& Phillips IAU$\,$231\footnote{Poster contributions at Symposium no. 231 of The International Astronomical Union, Asilomar 2005. Abstracts available at http://asilomar.caltech.edu/.}), and an IRAM 30 m survey
(168 GHz in three windows between 80 and 281 GHz) by
Tercero et al. (IAU$\,$231$^{1}$).

The first imaging line survey of Orion KL in the submm range (337.2--339.2 and
347.2--349.2 GHz) was recently reported by
Beuther et al. (\cite{beuther05}), who used the Submillimeter Array
interferometer.
They later employed the same instrument to make measurements
around 680 and 690 GHz with similar bandwidths
(Beuther at al. \cite{beuther06}).

The luminous Orion Kleinmann-Low infrared nebula
(Orion KL; $L\approx$10$^{5}\,L_{\sun}$), and its
surrounding molecular cloud, is the nearest (distance of 450 pc)
and probably most studied massive star formation region in the sky.
A very useful review has been written by
Genzel and Stutzki \cite{genzel89};
for reference updates see e.g., Olofsson et al. \cite{olofsson03},
and Wirstr\"om et al. \cite{wirstroem06}.
Here we summarise some
source component designations and dynamical properties particularly relevant
to the molecular line identification work in the current presentation of
our Odin spectral scan data, which includes our molecular line assignments
(in the Online Table~\ref{longtab}).

Odin's circa 126\arcsec\ antenna beam is centred on the most
prominent infrared `point' source in the KL nebula, IRc$\,$2 
(\radec{05}{35}{14}{36}{-05}{22}{29}{6}{J2000}).
The Orion \textit{hot core} source,
with a size of only $\approx$10\arcsec\ and centred only 2\arcsec\ S
of IRc$\,$2, is a warm ($\approx$200 K, or even higher;
cf. Sempere et al. \cite{sempere00}), dense ($\approx$10$^{7}$ cm$^{-3}$)
clump or rather collection of clumps, characterised by a spectral
line width of 5--15 \kmps\ centred on $v_{\rm LSR}$=3--6 \kmps\ and
exhibiting emission from nitrogen-containing species at markedly enhanced
abundances. The outflowing gas,
or the \textit{plateau source}, with a size of 40-60\arcsec, may be
characterised in terms of a bipolar \textit{high-velocity flow} elongated
in the SE-NW direction
(reaching velocities of $\pm$100 \kmps), and a SW-NE extended
\textit{low-velocity flow} (the `18 \kmps\ flow') of size 15--30\arcsec,
centred on 10 and 5 \kmps, respectively.
Further details on the complex structure of the outflow -- such as a central jet and localised `bullet' type emission within the high velocity flow -- are nicely revealed by the IRAM 30-m CO J=2-1 maps by Rodr\'i{}guez-Franco et al.~(\cite{rodriguez99}).
Within the Odin 126\arcsec\
antenna beam there is also an N-S extended quiescent molecular cloud structure
(the \textit{ridge}), with densities of 10$^{4}$--10$^{6}$ cm$^{-3}$\ and
temperatures in the range 20--60 K, and
characterised by line widths of 3--5 \kmps\ and an abrupt velocity shift
across the KL nebula from $v_{\rm LSR}$=8 \kmps\ (in the south) to
10 \kmps\ (in the north).
The large-scale chemical structure of many important ridge molecules is outlined in Ungerechts et al. (\cite{ungerechts97}).

The interaction between the bipolar high-velocity outflow and the
surrounding ridge gas produces shock heating and shock enhanced
chemistry, markedly visible in terms of bright H$_{2}$\ emission,
strong high-$J$\ CO lines and uniquely strong emission from abundant
H$_{2}$O (Melnick et al. \cite{melnick00};
Olofsson et al. \cite{olofsson03}). Also the roughly orthogonal low-velocity
outflow component appears to interact with the ambient gas, creating
density, temperature and column density enhancements in the ridge gas.
One such feature is the \textit{compact ridge cloud}
($\Delta{}v$=3 \kmps; $v_{\rm LSR}\approx$8 \kmps) situated only
10--15\arcsec\ S of IRc$\,$2, on the northern tip of the 8 \kmps\ ridge
cloud, where complex oxygen-containing molecules have been observed to be
abundant. The hot core itself might also be a result of shock-induced
compression. A recent finding along these lines is enhanced [C{\small{}I}]
forbidden line ($^{3}$P$_{2}$--$^{3}$P$_{1}$)
emission north and south of IRc$\,$2 (in a shell of radius
$\approx$20\arcsec, where the outflow encounters ambient gas),
and proposed to result from CO
dissociation in shocks (Pardo et al. \cite{pardo05}).

Although the major source constituents are commonly discussed in the
literature as outlined above, we caution readers that observations yielding
higher spatial resolutions
(e.g. Blake et al. \cite{blake96}; Wright et al. \cite{wright96};
Beuther et al. \cite{beuther05}, all employing aperture synthesis)
reveal that the KL core composition is in fact more complex and breaks
down into further sub-structures of sizes $\lesssim10^{3}$\ AU.

We present here the observational results and line
identifications. Numerical analyses such as column density and
rotation temperature estimates are included in an accompanying paper
in this A\&A issue
(Persson et al.~\cite{persson06}, Paper II hereafter).


\section{Observations}

The present data-set was obtained in a four-part campaign running
over 1.5 years, starting in spring 2004 (Feb.--Apr.), followed by fall
2004 (Aug.--Oct.) and continuing in the same manner in 2005.
This division naturally arises from the combination of source coordinates
and Odin's orbital plane (Sun-synchronous low Earth orbit) leading to
seasonal visiblilty constraints for low-declination sources such as the
Orion nebula.

The dedicated observing period spanned over $\sim$1100 revolutions with a very
high success rate due to consistently stable spacecraft performance.
Each orbit allows 61 minutes of astronomical observations, whereas
the source \mbox{line-of-sight} is occulted by the Earth and its atmosphere
for the remaining 35 minutes of the orbital period.
We employed position switching (PSW) in order to acquire cold sky
reference spectra. This was implemented by regularly reorienting the entire
spacecraft by $-$15\arcmin\ in RA with a cycle time of 1 minute,
carried out by the onboard momentum wheels.
The resulting efficiency penalty for slew time was low ($<$20\%).

The beam size and main beam efficiency of Odin's 1.1 m offset Gregorian
telescope at 557 GHz are 2\arcmin1 and 0.9, respectively, as measured in
continuum observations of Jupiter assuming a Jovian brightness temperature
of 145 K and a disc-like source geometry (Frisk et al. \cite{frisk03}).

The pointing is maintained in real time by the attitude control system,
assisted by two star trackers with an angular separation of 40\degr.
It has been empirically established that the reconstructed attitude
uncertainty is $\leq$15\arcsec most of the time.

There are four tunable submm receivers of single-sideband (SSB) type in the
radiometer, all of which were employed for these measurements.
Image band rejection is achieved using Martin-Puplett filters with cooled
termination absorbers and Schottky mixers are used for frequency
down-conversion.
The channels have centre frequencies of 495, 549, 555, and 572 GHz
and the design of the receiver system allows simultaneous use of
RX555 and RX572, or RX549 and RX495 (Frisk et al. \cite{frisk03}).

The tuning range of 14 GHz was not fully exploited in all channels although
the combined results from the complementary pair RX549 and RX555 cover all
their accessible frequency range.
System temperatures in the passband
centres were around 3000--3500 K with few exceptions as measured by switching
between the main beam and a hot load at room temperature.
For a given tuning, \Tsys\ does not change in time by more than a few percent
due to the stable conditions and no interfering atmosphere.

The observing strategy was to tune the receivers in 0.5 GHz steps
and observe for 25 orbits. The resulting overlap in between adjacent
tunings gives a net on-source integration time of
50$\times$61$\times$0.8/2$\approx$20h per channel. This commonly used
approach was adopted to reduce potential impacts of
artificial spectral patterns/transients or baseline effects which could
arise in one tuning but conceivably not in both.

Three spectrometers were used: one acousto-optical spectrometer (AOS) and
two hybrid autocorrelators (AC1/2). The former has a bandwidth of 1.1 GHz and
a channel spacing of 620 kHz, while the latter two can be used in several
modes. In the low resolution mode used here, the bandwidth is 700 MHz
with 1 MHz channel spacing.
The AOS and the ACs have rather different characteristics and
therefore we have used slightly different approaches in building
the final spectra, hence the division into two parts of the next section.

\section{Data reduction}

The data reduction has been performed in two parallell, independent
efforts by members of the Odin Team situated at the
Onsala Space Observatory and in the University of Calgary. The spectra
presented in this paper are the products of the data reduction at Onsala.
The very similar Calgary results have been used to verify the quality
of the spectra shown here. Although a few differences were found for lines
at a level below 0.1 K (or below 4$\sigma$\ in terms of the baseline noise),
we are encouraged to believe that the majority of such weak lines are real.

\subsection{AOS and data from RX549/RX555}

\subsubsection{Calibration and averaging}
\label{aoscalsec}
There were essentially three complicating factors that needed special
consideration in these data sets.
\begin{description}
 \item[\textbf{Frequency correction:}]
       By using telluric ozone and water (and occasionally rarer
       lines during the occultation phase, the capability
       exists to check the frequency setting to a very high
       accuracy, as is illustrated in Fig.~\ref{atmofig}.
\begin{figure}
\includegraphics{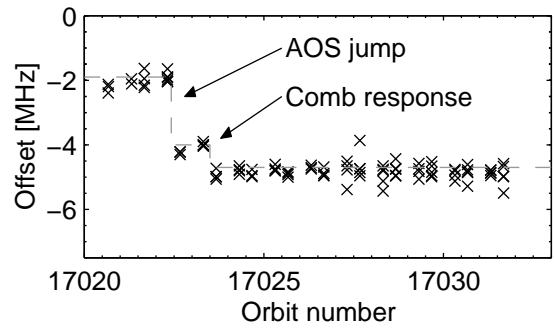}
\caption{Tracing the frequency deviation using the telluric H$_{2}^{17}$O line. As is evident, there was an early sudden shift of the line position, corresponding to an AOS laser mode jump during warming up -- a small penalty because of our shared astronomy/aeronomy mission. The times of such shifts can be determined accurately since they affect the total power levels of the spectrometer. The gray dashed line shows the correction applied to the data from this sample observation.}
\label{atmofig}
\end{figure}
       It was found that there was an offset of 0--5 MHz
       from the commanded value leaning towards the upper value
       most of the time. Furthermore, on a few occasions the offset
       was seen to change instantaneously by 2--3 MHz in mid-observation
       between two apparently stable offset levels.
       Thus it was deemed necessary to register telluric frequency references
       at all times and correct the astronomical spectra accordingly.
       If no atmospheric lines were present in a particular tuning,
       overlapping astronomical lines inherent to a different but
       adjacent tuning with a known frequency correction, were
       cross-correlated to find the proper correction. This method gave
       consistently positive results and we now believe that the absolute
       frequency uncertainty is 1 MHz ($\approx$0.5 \kmps) at worst.
 \item[\textbf{Channel dependent RMS:}]
       Probably owing to the large bandwidth of the AOS, the system
       temperature profiles had rather steep gradients towards the upper
       and lower edges, often leading to noise temperatures in these
       portions twice the mid-spectrum values. When averaging
       together two adjacent tunings, the low-noise central channels
       from one tuning will be combined with the high-noise edge channels
       of the other. This has been taken into account in the weighting
       procedure by letting the most recently measured \Tsys\ profile
       represent the formal noise \emph{in each channel} of the the next
       spectrum to be included in the average. Provided that the integration
       times and resolutions are all the same, and that we have a good
       linear correlation between \Tsys\ and the channel RMS, this approach
       is easily justified.
\item[\textbf{Side-band suppression:}]
       In general, we have very good suppression of the undesired image band
       ($\gtrsim$20 dB in at least one tuning). In the RX555 tunings,
       there is one exception where the appearance of an image SO
       line indicates a suppression of only 10 dB.
       Similarly poor SSB suppression was
       seen more frequently in the RX549 data through the emergence of
       telluric image ozone lines in occultation phase data. As is
       evident in the final result, we had in general more problems with
       high \Tsys\ and unexplained baseline ripples in this frontend/backend
       configuration, probably related to the poorer SSB performance.
       In those cases where image lines were seen in the final average
       (almost exclusively due to the strong \htwoo\ and \XIIIco\
       lines at 557 and 551 GHz, respectively), the appropriate channels were
       excluded to remove the interference. The complementary
       tunings certified that no gaps were created in the total average,
       although the noise is visibly higher in these portions due
       to less net observing time. Potential low-level image line intrusions
       are in general sufficiently weakened not to be seen in our
       baseline noise, owing to the satellite motion Doppler correction
       which serves to disperse image band signals.
\end{description}
Other pre-averaging measures consisted of first linearising the readouts
of the AOS aided by internal frequency comb measurements, then
resampling the AOS spectra to whole MHz to match the AC resolution.

The individual spectra each represent one on-source observation in the
PSW cycle which is equivalent to 24 s integration time. During one such
observation, the variation of the projected satellite orbital velocity in the
direction of the source is not taken into account which introduces a slight
spectral smearing of astronomical lines. However, the magnitude of the
Doppler shift change \emph{at most} amounts to
 $\sim$200 \mps\ and is considerbly less for the major part of the orbit.

The baseline stability was very good (Sect. \ref{contsubsec}) and no fits
were subtracted from the spectra in the averaging procedure.
However, to obtain accurate line property estimates for the very
weak features, we opted to subtract a piecewise linear fit from the
broadband end products, putting the baseline level at zero around clearly
detected emission lines.

\subsubsection{Baseline stability/continuum level}
\label{contsubsec}

\begin{figure}
  \centering
  \includegraphics[width=\columnwidth]{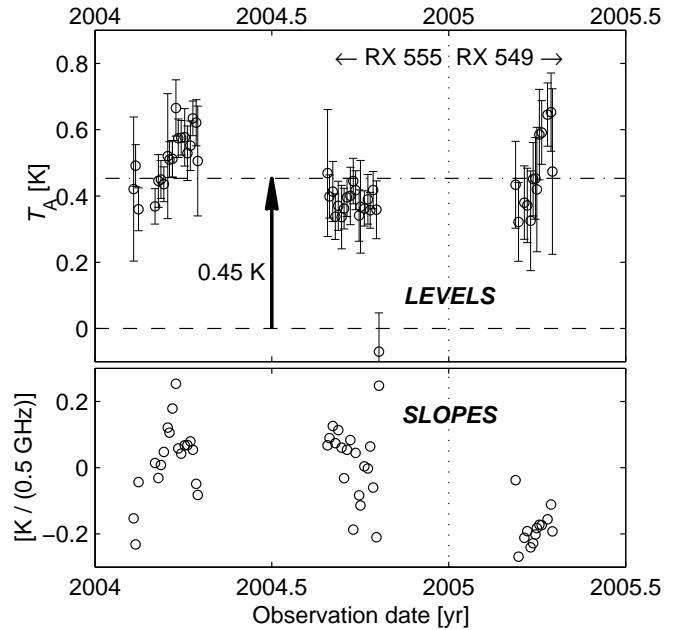}
  \caption{Baseline levels and slopes in 52 RX555 and RX549 observations before any baseline subtractions. The error bars in the upper panel correspond to 3$\sigma$. The slopes are measured as the level difference of two halves of the spectra whose centres are separated by 496 MHz. This figure indeed highlights the stability of the receiver/AOS/calibration chain, but the upwards trends of the levels towards the ends of both spring observing runs remain unexplained. There is also an observation that inexplicably appears to lack continuum signal (while telluric spectral lines could still be detected).}
  \label{contfig}
\end{figure}

We have estimated the dust continuum beam averaged antenna temperature
around 550 GHz using the spectra
from the RX549 and RX555 receivers. One value was extracted for each
tuning as shown in Fig. \ref{contfig}\ and the average amounted to
$T_{\rm A}$=0.45 K (1800 Jy$\,$beam$^{-1}$).

Due to the crowding of strong lines in some tunings, this calculation
relied on noise statistics and did \emph{not} require that the
positions of lines were known beforehand. The assumptions made were
instead i) at least half the channels in a spectrum were largely unaffected
by emission lines, ii) the baselines are largely flat compared to the
noise scatter of the intensity in individual spectra, and iii) the noise
is Gaussian and the RMS is well described by its formal value derived from
the radiometer formula.

The simple procedure was then to sort all channels according to increasing
intensity and select the bottom half of the distribution (the top half is
`contaminated' by emission lines). Statistically, the expectation value of
the distribution is then found by adding 0.8$\times$RMS$_{\rm formal}$\
to the mean value of the selected low-intensity channels.

To further remove effects of baseline variations, and to get a
handle on the baseline stability, each spectrum was divided into two
500 MHz subspectra on which the calculation described above were
performed. The average difference between the two halves in each tuning
then additionally supplied an indication of the first order trend
(or slope) of the baselines. Figure~\ref{contfig}\ also illustrates the
results found here. Although it is obvious that a systematic baseline
pattern was present throughout the RX549 observations, the RX555
baseline performance is quite satisfactory.

\subsection{ACs and data from RX495/RX572}

The 700 MHz AC spectra are created by stitching together seven 112 MHz
portions.
To achieve the double coverage of each frequency interval as we
do with the AOS, one AC was split into two portions, 3+4 subbands,
and placed on each side of the other AC that covered the central 700 MHz
in the passband.

Each such sub-band is typically ``well behaved'' in terms of having Gaussian
noise and a uniform gain curve.
However, in the PSW observing mode there is for unknown reasons
a linear falloff in channel intensity from the start of each on and off
sequence. Due to the inescapable slight asymmetry between on and
off measurements, and to the fact that the drift rate changes gradually over
the channels in one sub-band, the calibrated spectra end up with a low-level
curved saw tooth-like appearance with two sub-bands forming each tooth.
This pattern has been
removed by using high-order polynomials after careful comparisons
with the results obtained by calibrating each sub-band separately.
Around the extremely broad CO line at 576.3 GHz, this method failed and
the corresponding two tunings were instead cured by employing a corrective
procedure that nullifies the effects of the intensity drift mentioned above
but is very time-consuming.

The uppermost 200 MHz of the RX495 spectrum were acquired from a tuning
that suffered instability; the LO alternated between being
properly locked and oscillating between frequency offsets of $\pm$33 MHz.
Using the recorded IF current and the amplitudes of the [C{\small{}I}]
forbidden line line at 491.7 GHz (visible in individual spectra,
$T_{\rm A}${}$\approx$6 K) as guides, about ten minutes of integration
was recovered. In spite of the higher noise level thus obtained,
we chose to include this spectral section since it contains two
strong lines, the [C{\small{}I}] and a low-energy methanol transition
($T_{\rm A}${}$\approx$1.2 K).

We did not find any evidence of image band lines interfering in the
signal side band for these two receivers, nor did we find any significant
frequency offsets (at most $\approx{}1\,$MHz) from nominal values in
telluric line position controls in sample tunings.

\section{Results}

\subsection{Summary}
Table \ref{overtab} lists some general figures and characteristics for the
survey as a whole.
\begin{table}
\begin{minipage}[t]{\columnwidth}
\caption{Survey overview}             
\label{overtab}      
\centering
\renewcommand{\footnoterule}{}         
\begin{tabular}{l l} 
\hline\hline       
Property & Value \\
\hline                    
Bandwidth covered & 42 GHz \\
Number of lines detected\fn{Incl. U- and T-lines (see Sect.~\ref{ulinesec}).} & 344 \\  
Line density & 8 GHz$^{-1}$\ (range: 4--20) \\
Line intensity range & 0.05--80 K \\
Number of species detected\fn{Including isotopologues and \ion{C}{I}.} & 38 \\
Typical RMS reached & $\sim$25 mK/1 MHz channel \\
Total $\int\!{}T_{\mathrm{A}}\,{}dv$ & 2$\,$455\fn{Over all lines detected.}/2$\,$525\fn{Over all measured frequencies.}\ \Kkmps \\
Total line luminosity\fn{Using the expression 2.65$\times{}T_{\rm r}\times\theta/\lambda$\ where $\theta$=2.1\arcmin\ and $\lambda$\ is given in centimetres, to find the flux density per beam.}\fn{Using a beam-filling correction factor of 1.25 for data near 490 GHz.} & 5$\times$10$^{26}$\ W \\
Mean line-to-continuum ratio\fn{Near 550 GHz.} & 0.2 \\
\hline                  
\end{tabular}
\end{minipage}
\end{table}
As is evident from this Table, our integrated intensity is dominated
by the continuum emission by a ratio of 5:1 (see Sect. \ref{contsubsec}
for details).
The \mbox{line-to-continuum} ratio found here, 20\%, is much lower than that
at 350 GHz, 50\%, but rather close to the 15\% found at 650 GHz
(Schilke et al. \cite{schilke01}). For a larger
telescope (such as the Herschel Space Observatory) one can expect this value
to increase somewhat due to increased intensity from the abundant lines from
the compact line sources ($<$10\arcsec) compared to the dust continuum source
whose KL component is about 30\arcsec\ as seen in for example the 345 GHz
bolometer map of Siringo et al. (\cite{siringo04}).

\subsection{Spectra and identifications}

The final spectral scan results from the four receivers can be seen in
Figs. \ref{495fig}, \ref{549fig}, \ref{555fig}, and \ref{572fig}. The
frequency scale is counted w.r.t. a source velocity of $\vlsr$=$+$8 \kmps,
and markers are placed at the laboratory rest frequencies
of the transitions attributed to the line features.

The general method for identification was to select the most plausible
species after comparisons with available molecular
databases\footnote{On-line locations at\\
\url{http://www.cdms.de/}\hfill(CDMS)\\
\url{http://spec.jpl.nasa.gov/}\hfill(JPL)\\
(SLAIM03 is not available on-line but some of its content is maintained under \url{http://physics.nist.gov/PhysRefData/})}
(SLAIM03: Lovas \cite{lovas03}; CDMS: M\"uller et al. \cite{mueller01};
JPL: Pickett et al. \cite{pickett98}) of predicted/calculated or directly
measured transition frequencies (SLAIM03 contains both kinds wherever
available). The selection criteria included: frequency coincidence,
expected abundance, line strength, line width, line velocity,
and upper state energy. Where possible we have also used rotation diagram
analysis (cf. Goldsmith \& Langer \cite{goldsmith99}) to guide our
identification, as discussed in Paper II.

In the case of marginally detected line features suspected to arise from more
complex molecules such as dimethyl ether and methyl formate, we also required
that other lines of similar expected emission characteristics be
visible at other frequencies within the observed bands.

It is important to note that any conceivable artificial sharp features
produced in the radiometer would likely be stable in the sky frequency
rest frame and thus would be significantly smeared here since the
satellite motion Doppler correction of each spectrum varies
between -7 and +7 \kmps\ over one orbit.

The line counts, upper state energy ranges and total integrated intensities
for each detected molecule are listed in Table \ref{counttab}.
All identifications (molecule and laboratory rest frequency) can be found
in Table~\ref{longtab}\ (available on-line).

\input{tables/ctab}

Also of interest here are the spectral offsets of the measured lines relative
to the laboratory frequencies.
We have estimated these from the value found at the peak intensity channel
of the lines (assuming a systemic emission velocity of +8 \kmps) and they
are listed in full in Paper~II.
The average offset per molecule is $-$0.6 MHz but this cannot be used as a
quality measure of the data fidelity nor the accuracy of tabulated rest
frequencies since different molecules emit at different velocities (as is
likely reflected by the high dispersion, 3.2 MHz). In addition, these emission
velocities are often instrument-dependent due to varying beam fillings of
the different source components. 
Nonetheless, by choosing the 35 strongest lines of methanol (which conveniently
has narrow emission lines), we minimise this effect and get a dispersion of
only 1.1 MHz. While this is slightly higher than the stated estimated
frequency uncertainty of our spectra, there are a number of possible remaining
explanations aside from data error such as lab measurement/calculation error
and line blending.

\subsection{Important line blends}
A means to identify and subsequently remove `interfering' lines is
helpful, particularly in the detailed study of line shapes (where
the \mbox{signal-to-noise} ratios of the lines are such that this is feasible).

It now seems convincingly clear that the line wings of the rarer
isotopologues of
water and carbon monoxide are affected by emission from
SO$_{2}$, $^{34}$SO$_{2}$, CH$_{3}$OH and CH$_{3}$CN. Fortunately,
the emission from these species can be accurately modelled due to the
wealth of other lines from these species present in our band, allowing
determinations of column densities and excitation temperatures.
We have in the course of our analysis tried to reconstruct the true shapes
of some lines by subtracting polluting emissions from our spectra, and
in Paper~II one successful example is demonstrated (H$_{2}^{17}$O).

We caution that this approach is only useful in contexts where the
interfering lines emit in an optically thin portion of the main line
(or if the gas is stratified so that the optically thin line arises
in the near side of the gas column). For instance, we do not find
it likely that the SO$_{2}$\ line at 556.960 GHz significantly alters the
profile of the main water line (at 556.936 GHz) since i) the optical depth
towards the centre of the water line is very large, and ii) according to the
interpretation in Olofsson et al. (\cite{olofsson03}) the High Velocity Flow
as seen in water is located in front of (or around) the Low Velocity Flow
from which the sulphur oxide lines mainly originate, as evidenced by their
line widths and source size.

\subsection{Unidentified line features}
\label{ulinesec}
We have defined U-lines -- clearly detected lines well distinguished
from the baseline noise -- and T-lines, which are only marginally visible
(T for `Tentative') against the noise, or in an apparent line blend.
The lines of both types are marked in the spectra and listed in
Table \ref{uttab}.
\input{tables/uttab.tex}

We have in some cases found candidate species
(Table \ref{uttab}) which have not fulfilled all our criteria
for an unequivocal designation and they are kept as U- or T-lines.
Some of the more interesting scenarios are discussed below.

\subsubsection{New SO$^{+}$\ lines?}

Interstellar SO$^{+}$\ was first detected in the shocked clump IC443G,
presumably being formed in the dissociative shock caused by the
supernova remnant (Turner \cite{turner92}). Our weak unidentified
lines at 486.845 and 487.209 GHz tentatively can be identified as the
$J$=21/2--19/2 $e$\ and $f$\ doublet of the reactive radical SO$^{+}$\
in its $^{2}\Pi_{1/2}$\ ground state (JPL; Amano et al. \cite{amano91}).
The suggested assignment is consistent with the detection in Orion KL
of lower energy SO$^{+}$\ lines at 115.804, 116.180, 208.590, and 255.353 GHz
(Turner \cite{turner94}). Our assignment is further strengthened by the
U-line at 347.743 GHz observed by Schilke et al. (\cite{schilke97}) which
we identify here with the $J$=15/2--13/2 $e$\ transition of SO$^{+}$\
at 347.740 GHz. However, the corresponding $f$\ transition at 348.115 GHz
is hidden in a blend with $^{13}$CH$_{3}$OH and $^{34}$SO$_{2}$.

\subsubsection{Other notable frequency coincidences:\\ SH$^{-}$\ and ND}

Our weak, slightly broad lines near 546.138 and 546.176 GHz
could be associated with the $N_{J}$=1$_{1}$--0$_{1}$\ hyperfine
transition cluster of the ND radical in its $X^{3}\Sigma^{-}$\ ground
vibronic state
(CDMS; Saito \& Goto \cite{saito93}; Takano et al. \cite{takano98}).
If so, the strongest component is at a blue-shifted position closer to the
Hot Core velocity, in agreement to what is seen in other nitrogen hydrides
such as NH$_{2}$\ (which we tentatively conclude to be blue-shifted after
studying the 900 GHz spectral survey of Comito et al. \cite{comito05} in
detail). NH$_{3}$\ on the other hand -- discussed further in Paper~II -- has a
$\sim$60\% Hot Core contribution but this is heavily masked by its optical
depth in both the Hot Core and in the Compact Ridge. Unfortunately, the
predicted shape of the line bundle as a whole does not match the observation
and one would need to invoke non-LTE excitation to explain this difference.

A weak unidentified line at 564.418 GHz
may originate in the \jtrans{1}{0}\ rotational transition of
SH$^{-}$\ in its $X^{1}\Sigma^{+}$\ ground state, measured to fall at the
frequency 564.422 GHz (Civi\v{s} et al. \cite{civis98}).
A new interstellar anion would be of utmost interest since only one has
been found previously, namely the discovery of C$_{6}$H$^{-}$\ which was
recently reported by McCarthy et al. (\cite{mccarthy06}).
Thus -- although our line is not strong enough to claim a detection
($\sim$3.5$\sigma$\ relative to the local noise) -- we
chose to perform a very simple column density calculation (or an upper limit
thereof if the line turned out false). By employing the LTE assumption with
full beamfilling, an excitation temperature of 100 K, and a dipole moment
of 0.273 D (adopted from the ab initio calculations of
Senekowitsch et al. \cite{senekowitsch85}), we arrive at a figure of
$N_{\rm SH^{-}}$=2$\!\times\!$10$^{13}$\ \pscm.
The scenario used would be consistent with SH$^{-}$\ residing in the
extended OMC-1/M42 face-on PDR
(discussed in Wirstr\"om et al. \cite{wirstroem06}).
Should the origin of the line (if real) be any of the smaller Orion KL
components -- as indeed the observed line width
($\Delta{}v_{\rm FWHM}$=11 \kmps) seems to indicate -- the column
density rises by a factor of $\sim$200.
We deem at least the PDR column density to be a reasonably low value in light
of the rather unknown chemistry of similar hydrides in the interstellar medium.
The SH radical, for instance, has only been found in the atmosphere of a Mira
variable through mid-IR transitions (Yamamura et al. \cite{yamamura00}).
Nevertheless, a possible contribution from SH in a C$_{3}$\ spectrum
towards Sgr~B2 has been reported by Cernicharo et al. (\cite{cernicharo00}).
We also note that a formation pathway exists involving H$_{2}$S
(via dissociative electron attachment), a molecule observed by Odin in
the present survey and treated in Paper~II.

\vspace{.5ex}
Both these candidates (marked by gray arrows in the spectra figures)
can be confirmed or ruled out by the forthcoming Herschel mission
(discussed further in Sect.~\ref{discsec}). For the case of ND, the
triplets of NH at 947 and 1000 GHz are also relevant.

\subsection{Non-detections}
Two notable non-detections have been indicated by gray arrows in
our spectra: the PH and O$_2$\ molecules. The PH line (a hyperfine
transition within the $N_J$=1$_1$-0$_1$\ group at 553.363 GHz) has also been
searched for in IRC$+$10216 by Odin to a much higher sensitivity by
Bernath et al. (\cite{bernath07}), and we refer to them for
details on the PH physical and chemical properties and its role in the
circum- and interstellar medium. The O$_2$\ line at 487.249 GHz has
previously been searched for by the SWAS satellite (having a larger beam size)
down to a sensitivity of 6.4 mK (Goldsmith et al., \cite{goldsmith00})
without success.
Both O$_2$\ non-detections are consistent with the Odin upper limit from the
ground state transition at 118.750 GHz by Pagani et al. (\cite{pagani03}).

\subsection{Improvements of existing spectroscopic data}
\label{labfreqsec}

In the case of our rather weak H$_2$CS lines, we did at first find
frequency offsets between 6 and 14 MHz systematically red-wards of the
expected values (as given in the JPL database which indeed stated
high uncertainties for these calculated H$_2$CS frequencies). Prompted by
private communication, an entry for this molecule was subsequently
inserted into CDMS and the new calculated rest frequencies found there gave
very good agreements with our measured positions in all but one line.
This is an encouraging result proving that large measured frequency
offsets are likely not to be spurious and may in some cases indicate that the
existing tabulated spectroscopy data are off the mark.

In the case of H$_{2}$CS, new extensive laboratory spectroscopy work has
kindly been performed by Eric Herbst and his collaborators.
The resulting H$_{2}$CS database will be crucial for
Herschel Space Observatory.

In general, we hope that our measured line parameters (listed in bulk
in Paper II) may be useful for accurate determination of molecular data
such as rotational constants.

\section{Discussion of future observations}
\label{discsec}

No further Odin observations of this kind are currently planned and the
observation time cost for significant noise reduction would in any case be
prohibitive.
However, the spectral portions presented here will be reobserved by 
the Herschel space observatory\footnote{http://www.esa.int/science/herschel}
and is of interest to briefly discuss what they could obtain.
Herschel is a European Space Agency (ESA) satellite mission aimed at
launching a 3.5 m telescope equipped with very low noise \mbox{submm/far-IR}
heterodyne receivers in 2008.

Most of the identified lines in this survey belong to species already
observed at other transitions (at both lower and higher frequencies)
and source size estimates for the corresponding emission components are
available in the literature (e.g. references in Sect.~\ref{introsec}
and Paper~II). 
A rough Herschel beam-filling estimate based on these source sizes reveals
that those lines will be 1--10 times stronger in the Herschel spectrum,
with the majority leaning towards the higher end.
Assuming further an observation time of only one hour in combination with
recent figures for the Herschel receiver system sensitivity, one finds
that the signal-to-noise ratio will be increased by up to a factor of 20
in such a case.
This will be particularly useful for the weakest lines (5--10$\sigma$) seen
in the Odin survey which will be possible to study in some detail in the
Herschel spectrum. It is in this group we find nearly all the unidentified
lines and the potential for new discoveries among them is obvious.
The interesting frequency coincidences of SH$^{-}$\ and ND described
earlier are good examples.
The interpretation of these and even weaker lines runs, however, the risk of
being hampered by line crowding and blending (already up to
20 lines/GHz in our spectrum) due to the plethora of lines that will emerge
from the noise compared to the Odin spectrum. This in turn will put high
demands on the system baseline stability in order to correctly disentangle
the emissions.
The pollutive contributions in the key water isotopologue lines discussed
above will also be worsened (due to differential beam-fillings) and on
a side note one can predict that this problem will arise in most sources
where Herschel observes these species.

\section{Conclusions}

We have conducted a spectral survey in two submillimetre windows largely
inaccessible from the ground due to atmospheric opacity. The frequency
ranges covered are: 486.4--492.3 and 541.5--577.6 GHz. This was achieved
using the Odin submm satellite.

The spacecraft performance was generally excellent in terms of high
observing efficiency and good sideband suppression most of the time
(within the SSB receivers).

The baseline stability has been shown to be satisfactory, albeit
high-order polynomials being required in about 50\% of the data
to remove a fixed pattern arising in the autocorrelators in this
particular observation setup.

Careful attention has been paid to the frequency alignment of our
data resulting in an estimated frequency error of $\lesssim$1 MHz.

Thus, high confidence is warranted in the fidelity of the reduced spectra.
To ascertain the calibration accuracy, some lines have also been
compared to previous or later targeted Odin observations using markedly
different instrumental setups.

We found a total of 280 identified emission lines (some of which include
multiple transitions), and 28 unidentified lines. We have also pointed
out a further 36 borderline detected features which in some cases have
interesting candidate assignments such as SH$^{-}$\ and SO$^{+}$.

Among the 38 detected molecules, we have four water isotopologues seen
in at least five emssion lines. These data are used in Paper II to make
a water abundance analysis.

Column density estimates for all species are presented in Paper II,
as well as abundance, source size, and rotation temperature estimates
for a selection of molecules.

\begin{acknowledgements}
Generous financial support from the Research Councils and Space Agencies in Sweden, Canada, Finland and France is gratefully acknowledged. We sincerely thank Frank Lovas for a CD containing his molecular spectroscopy database SLAIM03, and the dedicated scientists at Cologne (CDMS) and at JPL for undertaking the all-important work of providing spectroscopic data through the internet.
\end{acknowledgements}

\appendix

\section{Spectra}

\begin{figure*}
  \centering
  \includegraphics[]{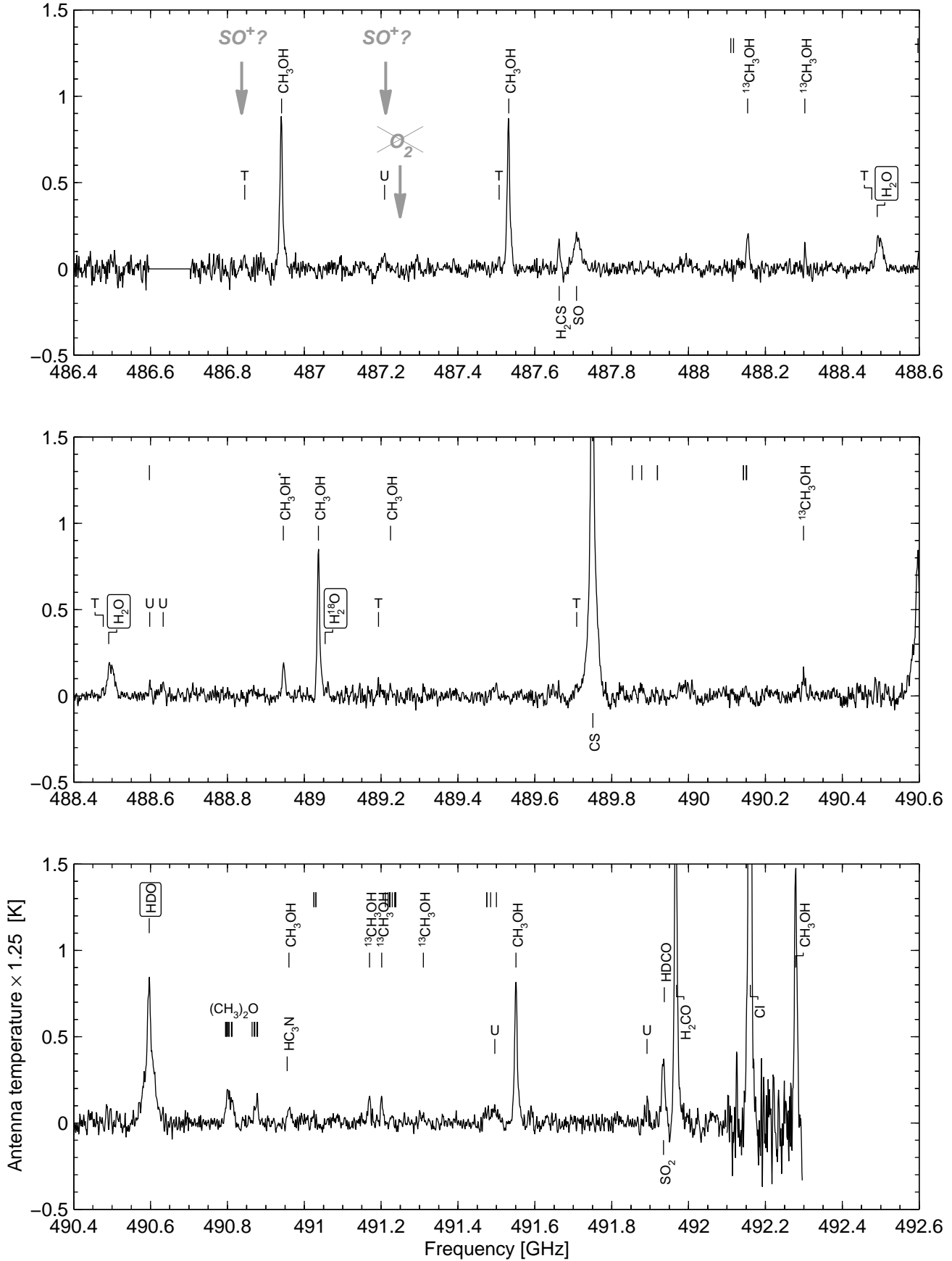}
  \caption{The Odin spectral survey between 486.4 and 492.1 GHz (RX495). Note the intensity scaling factor of 1.25 to approximately deconvolve the spectrum to the beam size near 557 GHz (assuming a point source). Empty markers at the intensity level of 1.25 K denote \metform\ lines.}
  \label{495fig}
\end{figure*}
\begin{figure*}
  \centering
  \includegraphics[]{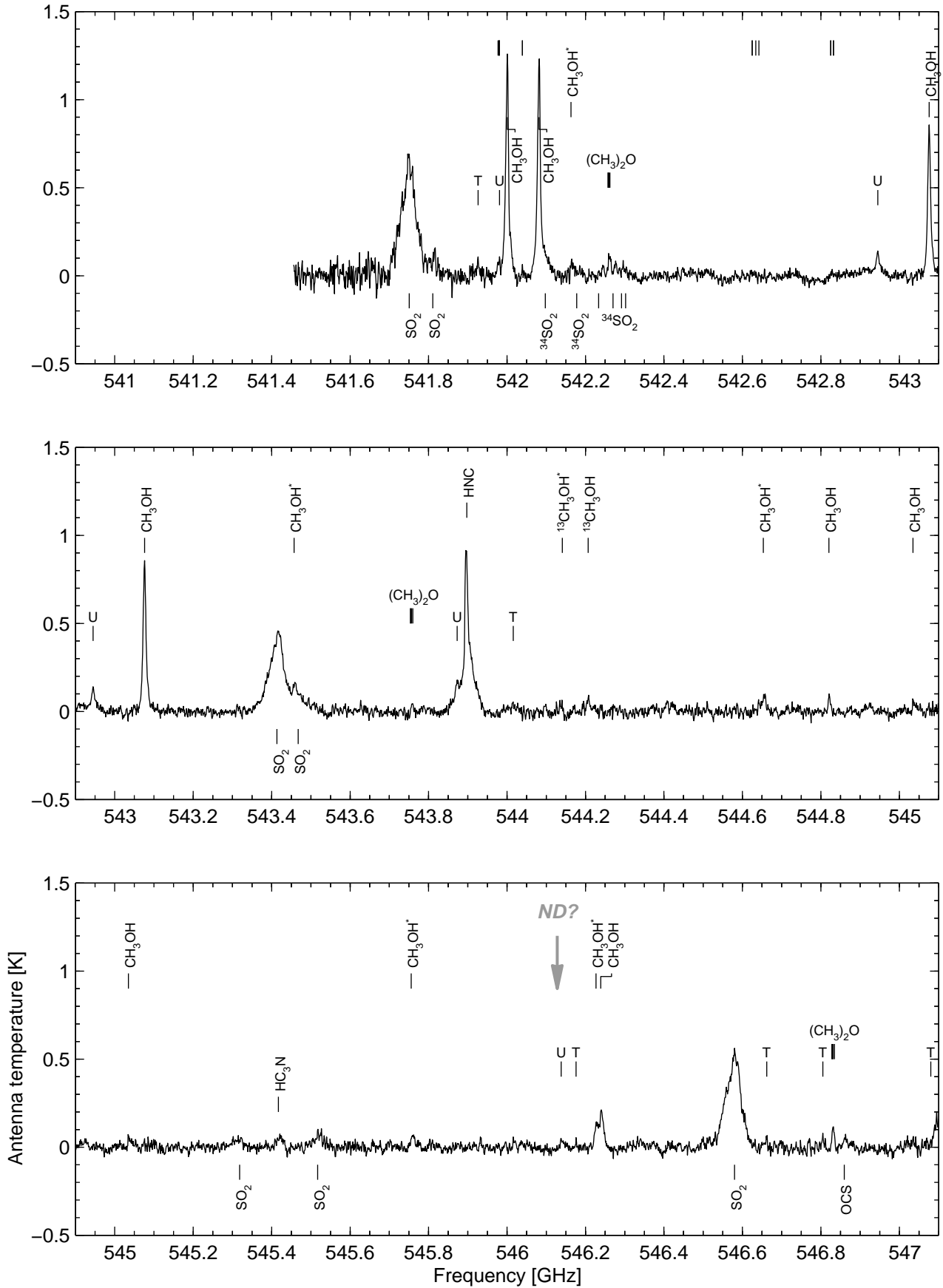}
  \caption{The Odin spectral survey between 541.4 and 547.1 GHz (RX549). Empty markers at the intensity level of 1.25 K denote \metform\ lines. A star denotes a transition within a vibrationally excited state: $v_{\rm t}$=1 for methanol, and $\nu_{\rm 2}$=1 for SO$_{2}$.}
  \label{549fig}
\end{figure*}
\begin{figure*}
  \centering
  \includegraphics[]{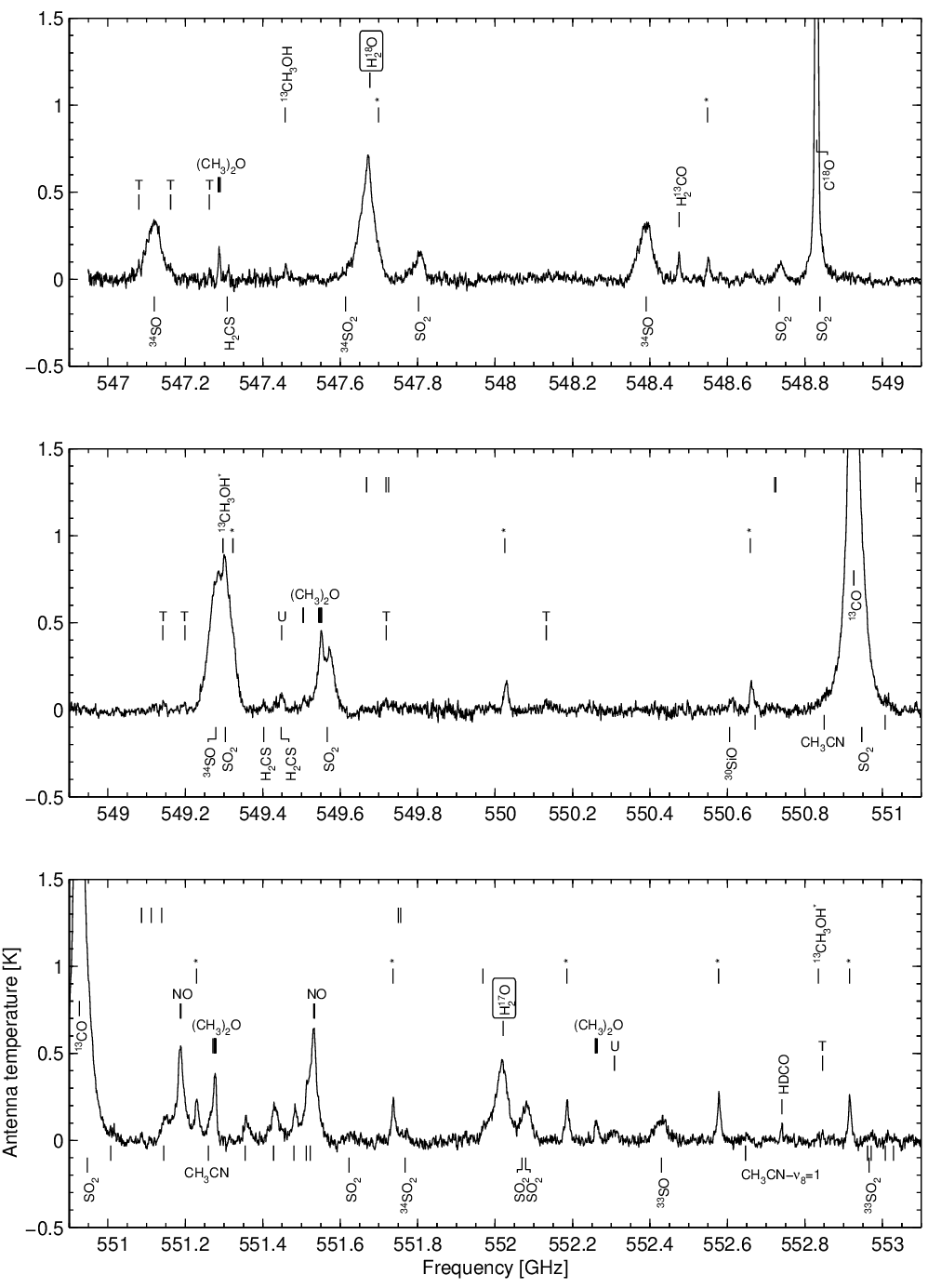}
  \caption{The Odin spectral survey between 546.9 and 563.1 GHz (RX555). Unlabelled markers at intensity levels -0.1, 0.5, 0.9, and 1.25 K, belong to \sotwo, \dimetet, \chIIIoh, and \metform, respectively. A star denotes a transition within a vibrationally excited state: $v_{\rm t}$=1 for methanol, and $\nu_{\rm 2}$=1 for SO$_{2}$. Two sets of markers immediately below the spectrum baseline belong to ground state and vibrationally excited \chIIIcn.}
  \label{555fig}
\end{figure*}
\begin{figure*}
  \centering
  \includegraphics[]{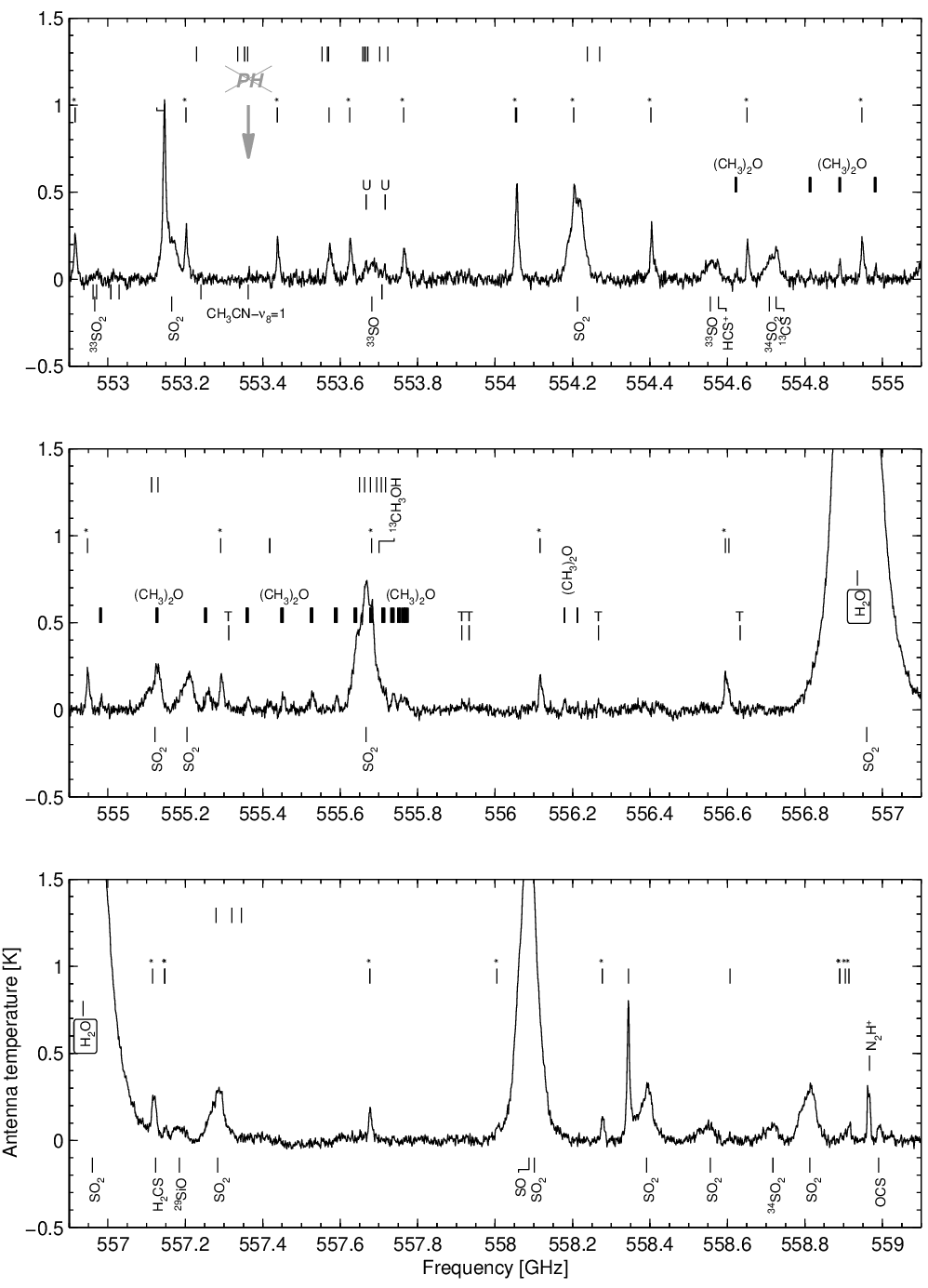}
  \addtocounter{figure}{-1}
  \caption{continued.}
\end{figure*}
\begin{figure*}
  \centering
  \includegraphics[]{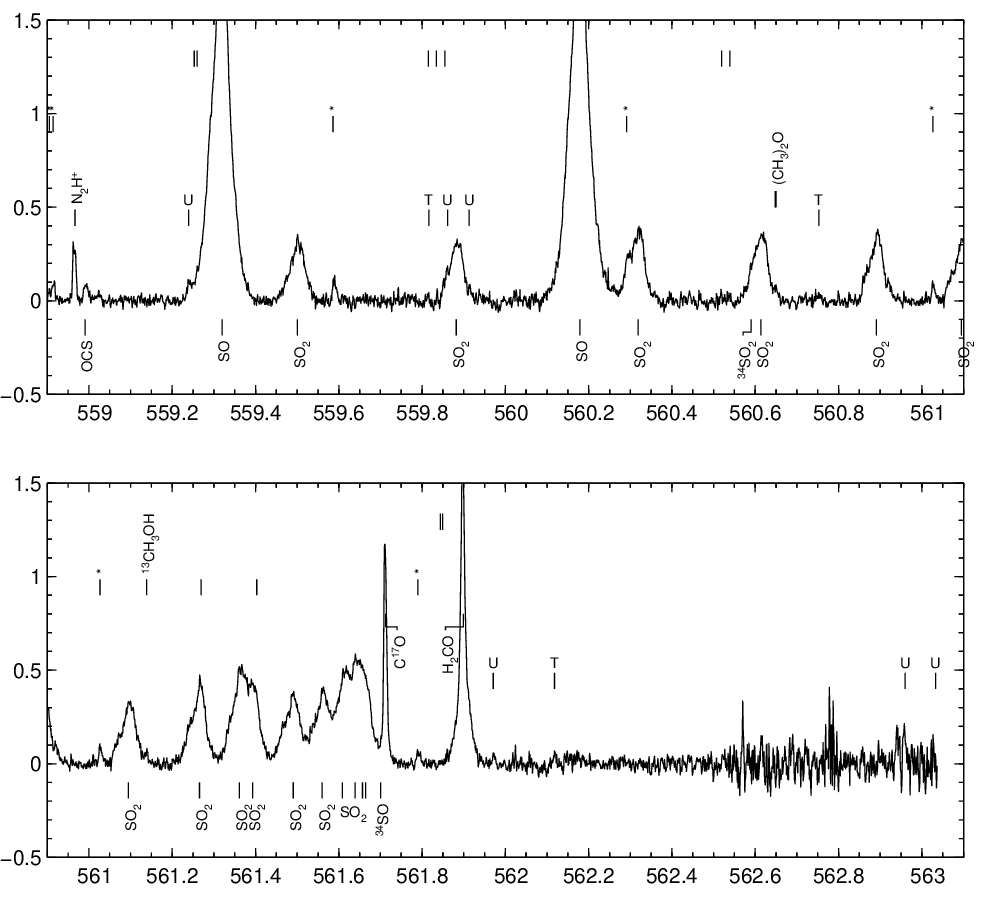}
  \addtocounter{figure}{-1}
  \caption{continued.}
\end{figure*}
\begin{figure*}
  \centering
  \includegraphics[]{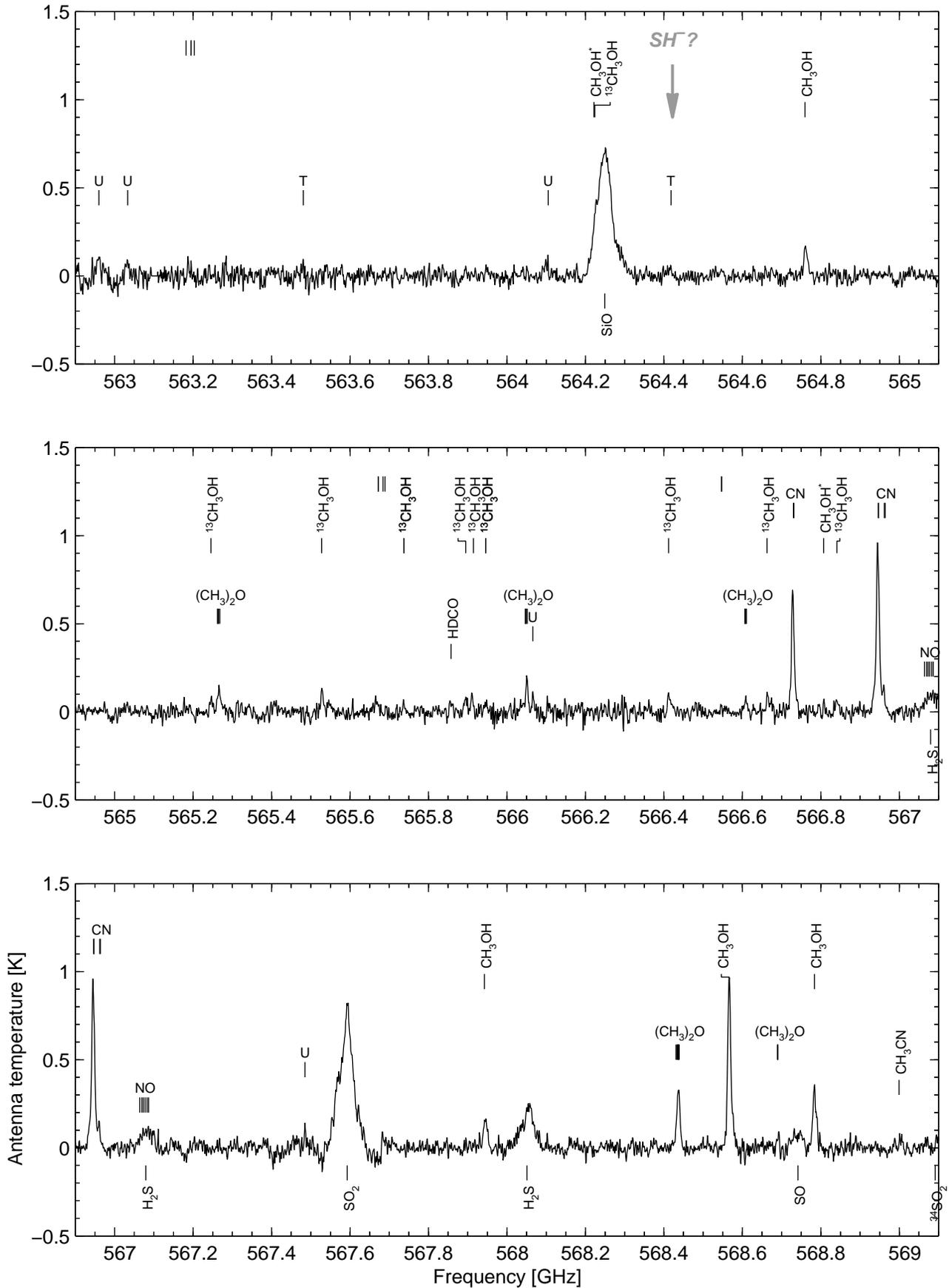}
  \caption{The Odin spectral survey between 562.9 and 577.6 GHz (RX572). Empty markers at the intensity level of 1.25 K denote \metform\ lines.}
  \label{572fig}
\end{figure*}
\begin{figure*}
  \centering
  \includegraphics[]{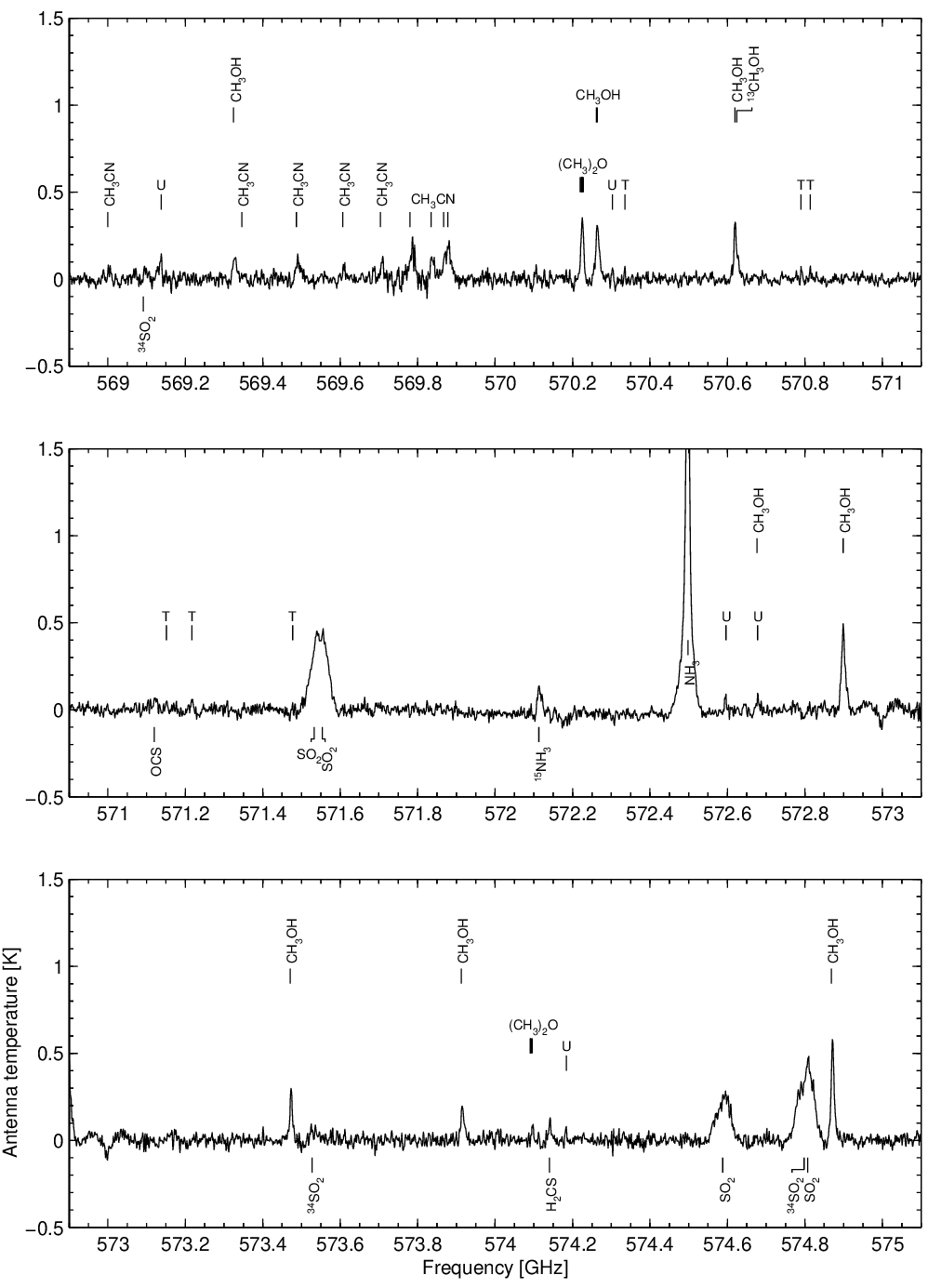}
  \addtocounter{figure}{-1}
  \caption{continued.}
\end{figure*}
\begin{figure*}
  \centering
  \includegraphics[]{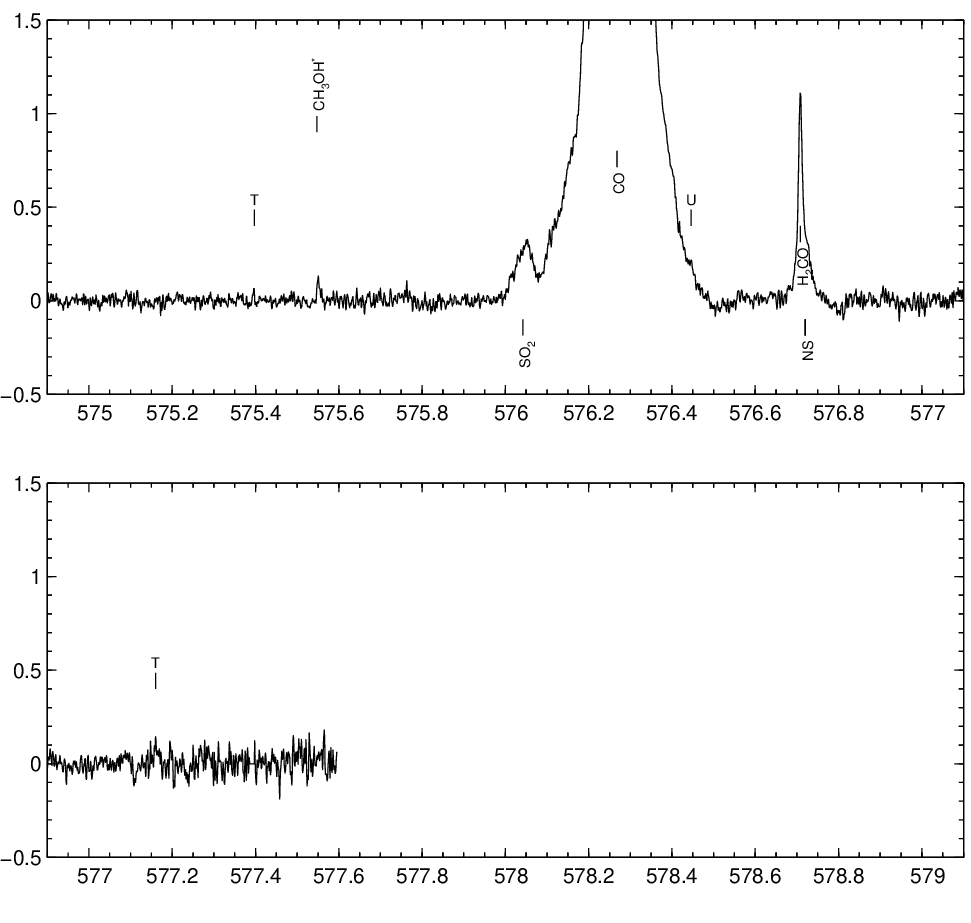}
  \addtocounter{figure}{-1}
  \caption{continued.}
\end{figure*}

\clearpage

\Online

\section{Electronic table of lines}

\input{tables/ltab}

\end{document}

%% file: tables/ctab.tex
\begin{table} 
\caption{Summary of  all detected species.} 
\label{counttab}
\begin{tabular} { l r r r}
\hline
\hline
Species  & Number & Energy range & $\int\!{}T_{\mathrm{A}}\,{}dv$ \\
	     &	            &	[K]              &  [\Kkmps{}]      \\                           
\hline
CH$_{3}$OCH$_{3}$		&	47	&	106-448	&	18.3\\ 
SO$_{2}$				&	42	&	75-737	&	239.9\\  
$^{34}$SO$_{2}$		&	5	&	79-457	&	2.6	\\  
SO					&	5	&	71-201	&	181.5\\  
$^{33}$SO			&	3	&	191-199	&	6.5\\ 
$^{34}$SO			&	2	&	191-197	&	15.6\\  
CH$_{3}$OH	$v_{\rm t}$=1	&	42	&	332-836	&		\\  
CH$_{3}$OH			&	34	&	38-863	&	108.6\\   
$^{13}$CH$_{3}$OH	&	21	&	37-499	&	6.8\\ 
$^{13}$CH$_{3}$OH $v_{\rm t}$=1	&	2	&	373-670	&		\\
CH$_{3}$CN			&	17	&	410-1012	&	9.3\\ 
NO					&	12	&	84-232	&	11.1\\  
CN					&	8	&	54		&	8.5\\  
H$_{2}$CS			&	5	&	138-343	&	1.7\\ 
H$_{2}$CO			&	3	&	106-133	&	38.5\\ 
H$_{2}^{13}$CO		&	1	&	130		&	0.6\\  
HDCO				&	3	&	114-141	&	2.8\\ 
OCS					&	3	&	604-658	&	1.3\\ 
$o$-H$_{2}$O			&	1	&	61	&		320.3\\ 
$p$-H$_{2}$O			&	1	&	867	&		2.2\\ 
$o$-H$_{2}^{17}$O		&	1	&	61		&	9.4\\ 
$o$-H$_{2}^{18}$O		&	2	&	60-430	&	16.2\\
HDO					&	1	&	66		&	10.7\\ 
HC$_{3}$N			&	2	&	648-799	&	0.6\\ 
CO					&	1	&	83		&	1100\\ 
$^{13}$CO			&	1	&	79		&	174.3\\ 
C$^{17}$O			&	1	&	81		&	6.0\\ 
C$^{18}$O			&	1	&	79		&	24.3\\ 
C					&	1	&	24		&	38.7\\ 
NH$_{3}$				&	1	&	27		&	20.6\\ 
$^{15}$NH$_{3}$		&	1	&	27		&	0.7\\ 
HNC					&	1	&	91		&	10.7\\ 
N$_{2}$H$^{+}$		&	1	&	94		&	1.5\\ 
H$_2$S				&	1	&	166		&	4.4\\ 
CS					&	1	&	129		&	24\\ 
$^{13}$CS                       &	1	&	173		& 	v blend\\ 
SiO					&	1	&	190		&	17.6\\ 
$^{29}$SiO			&	1	&	187		&	1.4\\ 
$^{30}$SiO			&	1	&	185		&	0.5\\
HCS$^{+}$			&	1	&	186		&	v blend \\ 
NS					&	1	&	442		&	v blend\\ 
U-line					&	28	&	-		&	9.1\\  
T-line				&	36	&	-		&	7.2\\  
\hline
\rem{
No species  	&		38	&\\
No lines 		&		344	&\\
Total $\int$ T$^{\,*}_{\mathrm{A}}$  d$v$ & &&		2\,455 \\
\hline   
}
\end{tabular}
\end{table}

%% file: tables/uttab.tex
\begin{table}
\caption{Unidentified (U) and/or marginally detected (T) lines}
\label{uttab}
\begin{tabular}{ccl|ccl}
\hline
\hline
Frequency & Type & Sugg. & \it{Table} & & \\
{}[MHz] & [U/T] & ident. & \it{continued} & & \hspace{6ex} \\
\hline
486845   & T & SO$^{+}$                   &     555914   & T & ---              \\
487209   & U & SO$^{+}$                   &     555933   & T & ---              \\
487507   & T & CH$_{3}$CHO                &     556267   & T & ---              \\
488477   & T & ---                        &     556633   & T & ---              \\
488598   & U & CH$_{3}$OCHO                &     559239   & U & CH$_{3}$OCHO     \\
488633   & U & CH$_{3}$CHO                &     559816   & T & HDO              \\
489193   & T & ---                        &     559861   & U & CH$_{3}$OCHO      \\
489709   & T & SiS                        &     559913   & U & ---              \\
491496   & U & CH$_{3}$OCHO               &     560753   & T & ---              \\
491892   & U & ---                        &     561971   & U & ---              \\
541926   & T & ---                        &     562118   & T & ---              \\
541981   & U & CH$_{3}$OCHO               &     562960   & U & ---              \\
542945   & U & ---                        &     563033   & U & ---              \\
543873   & U & ---                        &     563481   & T & HNCO             \\
544016   & T & SiS                        &     564105   & U & ---              \\
546138   & U & ND                         &     564418   & T & SH$^{-}$         \\
546176   & T & ND                         &     566066   & U & ---              \\
546662   & T & ---                        &     567485   & U & ---              \\
546805   & T & ---                        &     569138   & U & HNCO             \\
547080   & T & ---                        &     570303   & U & ---              \\
547162   & T & ---                        &     570335   & T & ---              \\
547262   & T & HNCO                       &     570790   & T & ---              \\
549142   & T & HNCO                       &     570814   & T & ---              \\
549199   & T & HNCO,H$_{2}$CS             &     571151   & T & HNCO             \\
549449   & U & ---                        &     571217   & T & HNCO             \\
549719   & T & SO$_{2}$,(CH$_{3}$)$_{2}$O &     571477   & T & H$_{2}$C$^{18}$O \\
550132   & T & SO$_{2}$                   &     572596   & U & ---              \\
552308   & U & ---                        &     572678   & U & ---              \\
552846   & T & ---                        &     574184   & U & ---              \\
553667   & U & CH$_{3}$OCHO                &     575397   & T & ---              \\
553716   & U & ---                        &     576446   & U & ---              \\
555312   & T & ---                        &     577160   & T & ---              \\
\hline
\end{tabular}
\end{table}

%% file: tables/ltab.tex
\begin{table*}
\label{longtab}
\caption{Lines detected towards Orion KL listed in order of laboratory frequency.}
\begin{minipage}[tl]{0.333\textwidth}
\begin{tabular}{ll}
\hline
\hline
Frequency & Molecule \\
{}[MHz] & or U/T-line \\
\hline
486845 & T-line \\
486940.9 & CH$_{3}$OH \\
487209 & U-line \\
487507 & T-line \\
487531.9 & CH$_{3}$OH \\
487663.4 & H$_{2}$CS \\
487708.5 & SO \\
488153.5 & $^{13}$CH$_{3}$OH \\
488302.6 & $^{13}$CH$_{3}$OH \\
488477 & T-line \\
488491.1 & $para$-H$_{2}$O \\
488598 & U-line \\
488633 & U-line \\
488945.5 & CH$_{3}$OH \\
489037.0 & CH$_{3}$OH \\
489054.3 & $ortho$-H$_{2}^{18}$O \\
489193 & T-line \\
489224.3 & CH$_{3}$OH \\
489709 & T-line \\
489751.1 & CS \\
490299.4 & $^{13}$CH$_{3}$OH \\
490596.7 & HDO \\
490795.3 & CH$_{3}$OCH$_{3}$ \\
490795.3 & CH$_{3}$OCH$_{3}$ \\
490797.5 & CH$_{3}$OCH$_{3}$ \\
490798.3 & CH$_{3}$OCH$_{3}$ \\
490801.3 & CH$_{3}$OCH$_{3}$ \\
490804.3 & CH$_{3}$OCH$_{3}$ \\
490804.7 & CH$_{3}$OCH$_{3}$ \\
490810.3 & CH$_{3}$OCH$_{3}$ \\
490811.4 & CH$_{3}$OCH$_{3}$ \\
490812.1 & CH$_{3}$OCH$_{3}$ \\
490864.4 & CH$_{3}$OCH$_{3}$ \\
490869.8 & CH$_{3}$OCH$_{3}$ \\
490871.5 & CH$_{3}$OCH$_{3}$ \\
490877.2 & CH$_{3}$OCH$_{3}$ \\
490877.6 & CH$_{3}$OCH$_{3}$ \\
490878.3 & CH$_{3}$OCH$_{3}$ \\
490955.7 & HC$_{3}$N \\
490960.0 & CH$_{3}$OH \\
491170.0 & $^{13}$CH$_{3}$OH \\
491201.3 & $^{13}$CH$_{3}$OH \\
491310.1 & $^{13}$CH$_{3}$OH \\
491496 & U-line \\
491550.9 & CH$_{3}$OH \\
491892 & U-line \\
491934.7 & SO$_{2}$ \\
491937.0 & HDCO \\
491968.4 & H$_{2}$CO \\
492160.7 & C \\
492278.7 & CH$_{3}$OH \\
541750.9 & SO$_{2}$ \\
541810.6 & SO$_{2}$ \\
\hline
\end{tabular}
\end{minipage}
\begin{minipage}[tc]{0.333\textwidth}
\begin{tabular}{ll}
\hline
\hline
Frequency & Molecule \\
{}[MHz] & or U/T-line \\
\hline
541926 & T-line \\
541981 & U-line \\
542000.9 & CH$_{3}$OH \\
542081.9 & CH$_{3}$OH \\
542097.7 & $^{34}$SO$_{2}$ \\
542163.0 & CH$_{3}$OH \\
542177.4 & $^{34}$SO$_{2}$ \\
542233.5 & $^{34}$SO$_{2}$ \\
542257.7 & CH$_{3}$OCH$_{3}$ \\
542257.7 & CH$_{3}$OCH$_{3}$ \\
542260.1 & CH$_{3}$OCH$_{3}$ \\
542262.4 & CH$_{3}$OCH$_{3}$ \\
542270.1 & $^{34}$SO$_{2}$ \\
542291.7 & $^{34}$SO$_{2}$ \\
542302.3 & $^{34}$SO$_{2}$ \\
542945 & U-line \\
543076.1 & CH$_{3}$OH \\
543413.5 & SO$_{2}$ \\
543457.3 & CH$_{3}$OH \\
543467.7 & SO$_{2}$ \\
543753.9 & CH$_{3}$OCH$_{3}$ \\
543753.9 & CH$_{3}$OCH$_{3}$ \\
543756.9 & CH$_{3}$OCH$_{3}$ \\
543759.8 & CH$_{3}$OCH$_{3}$ \\
543873 & U-line \\
543897.6 & HNC \\
544016 & T-line \\
544140.5 & $^{13}$CH$_{3}$OH \\
544206.7 & $^{13}$CH$_{3}$OH \\
544653.1 & CH$_{3}$OH \\
544820.5 & CH$_{3}$OH \\
545034.8 & CH$_{3}$OH \\
545318.5 & SO$_{2}$ \\
545417.1 & HC$_{3}$N \\
545517.3 & SO$_{2}$ \\
545755.6 & CH$_{3}$OH \\
546138 & U-line \\
546176 & T-line \\
546226.8 & CH$_{3}$OH \\
546239.0 & CH$_{3}$OH \\
546579.8 & SO$_{2}$ \\
546662 & T-line \\
546805 & T-line \\
546827.8 & CH$_{3}$OCH$_{3}$ \\
546829.2 & CH$_{3}$OCH$_{3}$ \\
546831.5 & CH$_{3}$OCH$_{3}$ \\
546834.5 & CH$_{3}$OCH$_{3}$ \\
546859.8 & OCS \\
547080 & T-line \\
547119.6 & $^{34}$SO \\
547162 & T-line \\
547262 & T-line \\
547284.8 & CH$_{3}$OCH$_{3}$ \\
\hline
\end{tabular}
\end{minipage}
\begin{minipage}[tr]{0.333\textwidth}
\begin{tabular}{ll}
\hline
\hline
Frequency & Molecule \\
{}[MHz] & or U/T-line \\
\hline
547286.2 & CH$_{3}$OCH$_{3}$ \\
547287.8 & CH$_{3}$OCH$_{3}$ \\
547290.1 & CH$_{3}$OCH$_{3}$ \\
547308.2 & H$_{2}$CS \\
547457.8 & $^{13}$CH$_{3}$OH \\
547613.5 & $^{34}$SO$_{2}$ \\
547676.4 & $ortho$-H$_{2}^{18}$O \\
547698.9 & CH$_{3}$OH \\
547802.2 & SO$_{2}$ \\
548389.8 & $^{34}$SO \\
548475.2 & H$_{2}^{13}$CO \\
548548.7 & CH$_{3}$OH \\
548734.3 & SO$_{2}$ \\
548831.0 & C$^{18}$O \\
548838.9 & SO$_{2}$ \\
549142 & T-line \\
549199 & T-line \\
549278.7 & $^{34}$SO \\
549297.1 & $^{13}$CH$_{3}$OH \\
549303.3 & SO$_{2}$ \\
549322.8 & CH$_{3}$OH \\
549402.4 & H$_{2}$CS \\
549447.5 & H$_{2}$CS \\
549449 & U-line \\
549504.7 & CH$_{3}$OCH$_{3}$ \\
549504.7 & CH$_{3}$OCH$_{3}$ \\
549504.9 & CH$_{3}$OCH$_{3}$ \\
549505.2 & CH$_{3}$OCH$_{3}$ \\
549543.8 & CH$_{3}$OCH$_{3}$ \\
549546.5 & CH$_{3}$OCH$_{3}$ \\
549547.4 & CH$_{3}$OCH$_{3}$ \\
549547.7 & CH$_{3}$OCH$_{3}$ \\
549550.4 & CH$_{3}$OCH$_{3}$ \\
549550.6 & CH$_{3}$OCH$_{3}$ \\
549550.8 & CH$_{3}$OCH$_{3}$ \\
549551.7 & CH$_{3}$OCH$_{3}$ \\
549566.4 & SO$_{2}$ \\
549719 & T-line \\
550025.2 & CH$_{3}$OH \\
550132 & T-line \\
550605.2 & $^{30}$SiO \\
550659.3 & CH$_{3}$OH \\
550671.7 & CH$_{3}$CN \\
550850.0 & CH$_{3}$CN \\
550926.3 & $^{13}$CO \\
550946.7 & SO$_{2}$ \\
551007.4 & CH$_{3}$CN \\
551143.9 & CH$_{3}$CN \\
551187.3 & NO \\
551187.5 & NO \\
551188.8 & NO \\
551228.6 & CH$_{3}$OH \\
551259.6 & CH$_{3}$CN \\
\hline
\end{tabular}
\end{minipage}
\end{table*}

\addtocounter{table}{-1}
\begin{table*}
\caption{Continued}
\begin{minipage}[tl]{0.333\textwidth}
\begin{tabular}{ll}
\hline
\hline
Frequency & Molecule \\
{}[MHz] & or U/T-line \\
\hline
551270.8 & CH$_{3}$OCH$_{3}$ \\
551273.6 & CH$_{3}$OCH$_{3}$ \\
551275.1 & CH$_{3}$OCH$_{3}$ \\
551275.1 & CH$_{3}$OCH$_{3}$ \\
551276.5 & CH$_{3}$OCH$_{3}$ \\
551276.5 & CH$_{3}$OCH$_{3}$ \\
551277.9 & CH$_{3}$OCH$_{3}$ \\
551279.4 & CH$_{3}$OCH$_{3}$ \\
551354.2 & CH$_{3}$CN \\
551427.9 & CH$_{3}$CN \\
551480.6 & CH$_{3}$CN \\
551512.2 & CH$_{3}$CN \\
551522.7 & CH$_{3}$CN \\
551531.5 & NO \\
551534.0 & NO \\
551534.1 & NO \\
551622.9 & SO$_{2}$ \\
551736.2 & CH$_{3}$OH \\
551767.4 & $^{34}$SO$_{2}$ \\
551968.8 & CH$_{3}$OH \\
552021.0 & $ortho$-H$_{2}^{17}$O \\
552069.4 & SO$_{2}$ \\
552078.9 & SO$_{2}$ \\
552184.8 & CH$_{3}$OH \\
552258.9 & CH$_{3}$OCH$_{3}$ \\
552258.9 & CH$_{3}$OCH$_{3}$ \\
552261.4 & CH$_{3}$OCH$_{3}$ \\
552264.0 & CH$_{3}$OCH$_{3}$ \\
552308 & U-line \\
552429.8 & $^{33}$SO \\
552577.2 & CH$_{3}$OH \\
552646.6 & CH$_{3}$CN \\
552740.9 & HDCO \\
552835.1 & $^{13}$CH$_{3}$OH \\
552846 & T-line \\
552915.4 & CH$_{3}$OH \\
552970.8 & CH$_{3}$CN \\
553007.8 & CH$_{3}$CN \\
553029.1 & CH$_{3}$CN \\
553146.3 & CH$_{3}$OH \\
553164.9 & SO$_{2}$ \\
553201.6 & CH$_{3}$OH \\
553240.3 & CH$_{3}$CN \\
553362.2 & CH$_{3}$CN \\
553437.5 & CH$_{3}$OH \\
553570.9 & CH$_{3}$OH \\
553624.5 & CH$_{3}$OH \\
553667 & U-line \\
553681.3 & $^{33}$SO \\
553707.6 & CH$_{3}$CN \\
553716 & U-line \\
553763.7 & CH$_{3}$OH \\
554052.7 & CH$_{3}$OH \\
\hline
\end{tabular}
\end{minipage}
\begin{minipage}[tc]{0.333\textwidth}
\begin{tabular}{ll}
\hline
\hline
Frequency & Molecule \\
{}[MHz] & or U/T-line \\
\hline
554055.5 & CH$_{3}$OH \\
554202.9 & CH$_{3}$OH \\
554212.8 & SO$_{2}$ \\
554402.5 & CH$_{3}$OH \\
554555.6 & $^{33}$SO \\
554576.6 & HCS$^{+}$ \\
554619.8 & CH$_{3}$OCH$_{3}$ \\
554619.8 & CH$_{3}$OCH$_{3}$ \\
554621.0 & CH$_{3}$OCH$_{3}$ \\
554621.5 & CH$_{3}$OCH$_{3}$ \\
554622.1 & CH$_{3}$OCH$_{3}$ \\
554622.6 & CH$_{3}$OCH$_{3}$ \\
554622.6 & CH$_{3}$OCH$_{3}$ \\
554623.2 & CH$_{3}$OCH$_{3}$ \\
554650.9 & CH$_{3}$OH \\
554708.2 & $^{34}$SO$_{2}$ \\
554726.0 & $^{13}$CS \\
554811.5 & CH$_{3}$OCH$_{3}$ \\
554811.5 & CH$_{3}$OCH$_{3}$ \\
554812.7 & CH$_{3}$OCH$_{3}$ \\
554813.4 & CH$_{3}$OCH$_{3}$ \\
554813.9 & CH$_{3}$OCH$_{3}$ \\
554814.6 & CH$_{3}$OCH$_{3}$ \\
554814.6 & CH$_{3}$OCH$_{3}$ \\
554815.3 & CH$_{3}$OCH$_{3}$ \\
554888.3 & CH$_{3}$OCH$_{3}$ \\
554888.3 & CH$_{3}$OCH$_{3}$ \\
554890.4 & CH$_{3}$OCH$_{3}$ \\
554892.5 & CH$_{3}$OCH$_{3}$ \\
554947.4 & CH$_{3}$OH \\
554979.1 & CH$_{3}$OCH$_{3}$ \\
554979.1 & CH$_{3}$OCH$_{3}$ \\
554980.4 & CH$_{3}$OCH$_{3}$ \\
554981.2 & CH$_{3}$OCH$_{3}$ \\
554981.7 & CH$_{3}$OCH$_{3}$ \\
554982.5 & CH$_{3}$OCH$_{3}$ \\
554982.5 & CH$_{3}$OCH$_{3}$ \\
554983.3 & CH$_{3}$OCH$_{3}$ \\
555121.5 & SO$_{2}$ \\
555124.5 & CH$_{3}$OCH$_{3}$ \\
555124.5 & CH$_{3}$OCH$_{3}$ \\
555125.9 & CH$_{3}$OCH$_{3}$ \\
555126.8 & CH$_{3}$OCH$_{3}$ \\
555127.3 & CH$_{3}$OCH$_{3}$ \\
555128.2 & CH$_{3}$OCH$_{3}$ \\
555128.2 & CH$_{3}$OCH$_{3}$ \\
555129.1 & CH$_{3}$OCH$_{3}$ \\
555204.1 & SO$_{2}$ \\
555249.8 & CH$_{3}$OCH$_{3}$ \\
555249.8 & CH$_{3}$OCH$_{3}$ \\
555251.3 & CH$_{3}$OCH$_{3}$ \\
555252.3 & CH$_{3}$OCH$_{3}$ \\
555252.7 & CH$_{3}$OCH$_{3}$ \\
\hline
\end{tabular}
\end{minipage}
\begin{minipage}[tr]{0.333\textwidth}
\begin{tabular}{ll}
\hline
\hline
Frequency & Molecule \\
{}[MHz] & or U/T-line \\
\hline
555253.8 & CH$_{3}$OCH$_{3}$ \\
555253.8 & CH$_{3}$OCH$_{3}$ \\
555254.9 & CH$_{3}$OCH$_{3}$ \\
555291.1 & CH$_{3}$OH \\
555312 & T-line \\
555356.8 & CH$_{3}$OCH$_{3}$ \\
555356.8 & CH$_{3}$OCH$_{3}$ \\
555358.3 & CH$_{3}$OCH$_{3}$ \\
555359.5 & CH$_{3}$OCH$_{3}$ \\
555359.9 & CH$_{3}$OCH$_{3}$ \\
555361.1 & CH$_{3}$OCH$_{3}$ \\
555361.1 & CH$_{3}$OCH$_{3}$ \\
555362.3 & CH$_{3}$OCH$_{3}$ \\
555417.7 & CH$_{3}$OH \\
555418.5 & CH$_{3}$OH \\
555447.3 & CH$_{3}$OCH$_{3}$ \\
555447.3 & CH$_{3}$OCH$_{3}$ \\
555448.9 & CH$_{3}$OCH$_{3}$ \\
555450.2 & CH$_{3}$OCH$_{3}$ \\
555450.6 & CH$_{3}$OCH$_{3}$ \\
555451.9 & CH$_{3}$OCH$_{3}$ \\
555451.9 & CH$_{3}$OCH$_{3}$ \\
555453.1 & CH$_{3}$OCH$_{3}$ \\
555522.9 & CH$_{3}$OCH$_{3}$ \\
555522.9 & CH$_{3}$OCH$_{3}$ \\
555524.6 & CH$_{3}$OCH$_{3}$ \\
555526.0 & CH$_{3}$OCH$_{3}$ \\
555526.4 & CH$_{3}$OCH$_{3}$ \\
555527.8 & CH$_{3}$OCH$_{3}$ \\
555527.8 & CH$_{3}$OCH$_{3}$ \\
555529.2 & CH$_{3}$OCH$_{3}$ \\
555585.3 & CH$_{3}$OCH$_{3}$ \\
555585.3 & CH$_{3}$OCH$_{3}$ \\
555587.1 & CH$_{3}$OCH$_{3}$ \\
555588.6 & CH$_{3}$OCH$_{3}$ \\
555589.0 & CH$_{3}$OCH$_{3}$ \\
555590.4 & CH$_{3}$OCH$_{3}$ \\
555590.4 & CH$_{3}$OCH$_{3}$ \\
555591.9 & CH$_{3}$OCH$_{3}$ \\
555635.9 & CH$_{3}$OCH$_{3}$ \\
555635.9 & CH$_{3}$OCH$_{3}$ \\
555637.9 & CH$_{3}$OCH$_{3}$ \\
555639.4 & CH$_{3}$OCH$_{3}$ \\
555639.8 & CH$_{3}$OCH$_{3}$ \\
555641.4 & CH$_{3}$OCH$_{3}$ \\
555641.4 & CH$_{3}$OCH$_{3}$ \\
555642.9 & CH$_{3}$OCH$_{3}$ \\
555666.3 & SO$_{2}$ \\
555676.3 & CH$_{3}$OCH$_{3}$ \\
555676.3 & CH$_{3}$OCH$_{3}$ \\
555678.3 & CH$_{3}$OCH$_{3}$ \\
555679.9 & CH$_{3}$OCH$_{3}$ \\
555680.3 & CH$_{3}$OCH$_{3}$ \\
\hline
\end{tabular}
\end{minipage}
\end{table*}

\addtocounter{table}{-1}
\begin{table*}
\caption{Continued}
\begin{minipage}[tl]{0.333\textwidth}
\begin{tabular}{ll}
\hline
\hline
Frequency & Molecule \\
{}[MHz] & or U/T-line \\
\hline
555681.0 & CH$_{3}$OH \\
555681.9 & CH$_{3}$OCH$_{3}$ \\
555681.9 & CH$_{3}$OCH$_{3}$ \\
555683.6 & CH$_{3}$OCH$_{3}$ \\
555700.1 & $^{13}$CH$_{3}$OH \\
555707.6 & CH$_{3}$OCH$_{3}$ \\
555707.6 & CH$_{3}$OCH$_{3}$ \\
555709.7 & CH$_{3}$OCH$_{3}$ \\
555711.4 & CH$_{3}$OCH$_{3}$ \\
555711.8 & CH$_{3}$OCH$_{3}$ \\
555713.5 & CH$_{3}$OCH$_{3}$ \\
555713.5 & CH$_{3}$OCH$_{3}$ \\
555715.2 & CH$_{3}$OCH$_{3}$ \\
555731.2 & CH$_{3}$OCH$_{3}$ \\
555731.2 & CH$_{3}$OCH$_{3}$ \\
555733.4 & CH$_{3}$OCH$_{3}$ \\
555735.2 & CH$_{3}$OCH$_{3}$ \\
555735.5 & CH$_{3}$OCH$_{3}$ \\
555737.3 & CH$_{3}$OCH$_{3}$ \\
555737.3 & CH$_{3}$OCH$_{3}$ \\
555739.1 & CH$_{3}$OCH$_{3}$ \\
555748.2 & CH$_{3}$OCH$_{3}$ \\
555748.2 & CH$_{3}$OCH$_{3}$ \\
555750.5 & CH$_{3}$OCH$_{3}$ \\
555752.3 & CH$_{3}$OCH$_{3}$ \\
555752.7 & CH$_{3}$OCH$_{3}$ \\
555754.6 & CH$_{3}$OCH$_{3}$ \\
555754.6 & CH$_{3}$OCH$_{3}$ \\
555756.4 & CH$_{3}$OCH$_{3}$ \\
555759.7 & CH$_{3}$OCH$_{3}$ \\
555759.7 & CH$_{3}$OCH$_{3}$ \\
555762.1 & CH$_{3}$OCH$_{3}$ \\
555764.0 & CH$_{3}$OCH$_{3}$ \\
555764.4 & CH$_{3}$OCH$_{3}$ \\
555766.3 & CH$_{3}$OCH$_{3}$ \\
555766.3 & CH$_{3}$OCH$_{3}$ \\
555766.8 & CH$_{3}$OCH$_{3}$ \\
555766.8 & CH$_{3}$OCH$_{3}$ \\
555768.2 & CH$_{3}$OCH$_{3}$ \\
555769.2 & CH$_{3}$OCH$_{3}$ \\
555771.2 & CH$_{3}$OCH$_{3}$ \\
555771.6 & CH$_{3}$OCH$_{3}$ \\
555773.5 & CH$_{3}$OCH$_{3}$ \\
555773.5 & CH$_{3}$OCH$_{3}$ \\
555775.5 & CH$_{3}$OCH$_{3}$ \\
555914 & T-line \\
555933 & T-line \\
556115.8 & CH$_{3}$OH \\
556179.4 & CH$_{3}$OCH$_{3}$ \\
556179.4 & CH$_{3}$OCH$_{3}$ \\
556179.4 & CH$_{3}$OCH$_{3}$ \\
556179.5 & CH$_{3}$OCH$_{3}$ \\
556212.0 & CH$_{3}$OCH$_{3}$ \\
\hline
\end{tabular}
\end{minipage}
\begin{minipage}[tc]{0.333\textwidth}
\begin{tabular}{ll}
\hline
\hline
Frequency & Molecule \\
{}[MHz] & or U/T-line \\
\hline
556212.0 & CH$_{3}$OCH$_{3}$ \\
556212.1 & CH$_{3}$OCH$_{3}$ \\
556212.2 & CH$_{3}$OCH$_{3}$ \\
556267 & T-line \\
556594.4 & CH$_{3}$OH \\
556603.7 & CH$_{3}$OH \\
556633 & T-line \\
556936.0 & $ortho$-H$_{2}$O \\
556959.9 & SO$_{2}$ \\
557115.2 & CH$_{3}$OH \\
557123.2 & H$_{2}$CS \\
557147.0 & CH$_{3}$OH \\
557147.0 & CH$_{3}$OH \\
557184.4 & $^{29}$SiO \\
557283.2 & SO$_{2}$ \\
557676.6 & CH$_{3}$OH \\
558004.6 & CH$_{3}$OH \\
558087.7 & SO \\
558101.2 & SO$_{2}$ \\
558276.8 & CH$_{3}$OH \\
558344.5 & CH$_{3}$OH \\
558390.9 & SO$_{2}$ \\
558555.8 & SO$_{2}$ \\
558606.4 & CH$_{3}$OH \\
558717.5 & $^{34}$SO$_{2}$ \\
558812.5 & SO$_{2}$ \\
558890.2 & CH$_{3}$OH \\
558890.2 & CH$_{3}$OH \\
558904.6 & CH$_{3}$OH \\
558914.0 & CH$_{3}$OH \\
558966.6 & N$_{2}$H$^{+}$ \\
558990.5 & OCS \\
559239 & U-line \\
559319.8 & SO \\
559500.4 & SO$_{2}$ \\
559586.1 & CH$_{3}$OH \\
559816 & T-line \\
559861 & U-line \\
559882.1 & SO$_{2}$ \\
559913 & U-line \\
560178.7 & SO \\
560291.0 & CH$_{3}$OH \\
560318.9 & SO$_{2}$ \\
560590.2 & $^{34}$SO$_{2}$ \\
560613.5 & SO$_{2}$ \\
560648.9 & CH$_{3}$OCH$_{3}$ \\
560648.9 & CH$_{3}$OCH$_{3}$ \\
560649.0 & CH$_{3}$OCH$_{3}$ \\
560649.0 & CH$_{3}$OCH$_{3}$ \\
560753 & T-line \\
560891.0 & SO$_{2}$ \\
561026.4 & CH$_{3}$OH \\
561094.8 & SO$_{2}$ \\
\hline
\end{tabular}
\end{minipage}
\begin{minipage}[tr]{0.333\textwidth}
\begin{tabular}{ll}
\hline
\hline
Frequency & Molecule \\
{}[MHz] & or U/T-line \\
\hline
561138.5 & $^{13}$CH$_{3}$OH \\
561265.6 & SO$_{2}$ \\
561269.1 & CH$_{3}$OH \\
561361.4 & SO$_{2}$ \\
561392.9 & SO$_{2}$ \\
561402.9 & CH$_{3}$OH \\
561490.5 & SO$_{2}$ \\
561560.3 & SO$_{2}$ \\
561608.6 & SO$_{2}$ \\
561639.3 & SO$_{2}$ \\
561656.7 & SO$_{2}$ \\
561664.2 & SO$_{2}$ \\
561700.4 & $^{34}$SO \\
561712.8 & C$^{17}$O \\
561789.7 & CH$_{3}$OH \\
561899.3 & H$_{2}$CO \\
561971 & U-line \\
562118 & T-line \\
562960 & U-line \\
563033 & U-line \\
563481 & T-line \\
564105 & U-line \\
564221.6 & CH$_{3}$OH \\
564223.7 & $^{13}$CH$_{3}$OH \\
564249.2 & SiO \\
564418 & T-line \\
564759.8 & CH$_{3}$OH \\
565245.2 & $^{13}$CH$_{3}$OH \\
565262.1 & CH$_{3}$OCH$_{3}$ \\
565262.8 & CH$_{3}$OCH$_{3}$ \\
565265.2 & CH$_{3}$OCH$_{3}$ \\
565267.9 & CH$_{3}$OCH$_{3}$ \\
565527.8 & $^{13}$CH$_{3}$OH \\
565737.4 & $^{13}$CH$_{3}$OH \\
565737.4 & $^{13}$CH$_{3}$OH \\
565857.5 & HDCO \\
565895.0 & $^{13}$CH$_{3}$OH \\
565914.4 & $^{13}$CH$_{3}$OH \\
565946.2 & $^{13}$CH$_{3}$OH \\
566046.6 & CH$_{3}$OCH$_{3}$ \\
566047.3 & CH$_{3}$OCH$_{3}$ \\
566049.3 & CH$_{3}$OCH$_{3}$ \\
566051.7 & CH$_{3}$OCH$_{3}$ \\
566066 & U-line \\
566411.9 & $^{13}$CH$_{3}$OH \\
566606.7 & CH$_{3}$OCH$_{3}$ \\
566606.7 & CH$_{3}$OCH$_{3}$ \\
566608.5 & CH$_{3}$OCH$_{3}$ \\
566610.4 & CH$_{3}$OCH$_{3}$ \\
566662.8 & $^{13}$CH$_{3}$OH \\
566729.9 & CN \\
566730.7 & CN \\
566730.8 & CN \\
\hline
\end{tabular}
\end{minipage}
\end{table*}

\addtocounter{table}{-1}
\begin{table*}
\caption{Continued}
\begin{minipage}[tl]{0.333\textwidth}
\begin{tabular}{ll}
\hline
\hline
Frequency & Molecule \\
{}[MHz] & or U/T-line \\
\hline
566840.7 & $^{13}$CH$_{3}$OH \\
566946.8 & CN \\
566946.9 & CN \\
566947.2 & CN \\
566962.0 & CN \\
566963.7 & CN \\
567064.2 & NO \\
567069.6 & NO \\
567073.4 & NO \\
567077.9 & NO \\
567079.6 & H$_{2}$S \\
567082.7 & NO \\
567086.6 & NO \\
567485 & U-line \\
567592.7 & SO$_{2}$ \\
567942.6 & CH$_{3}$OH \\
568050.7 & H$_{2}$S \\
568430.6 & CH$_{3}$OCH$_{3}$ \\
568432.6 & CH$_{3}$OCH$_{3}$ \\
568433.9 & CH$_{3}$OCH$_{3}$ \\
568435.3 & CH$_{3}$OCH$_{3}$ \\
568436.3 & CH$_{3}$OCH$_{3}$ \\
568437.2 & CH$_{3}$OCH$_{3}$ \\
568437.7 & CH$_{3}$OCH$_{3}$ \\
568439.2 & CH$_{3}$OCH$_{3}$ \\
568566.1 & CH$_{3}$OH \\
568690.1 & CH$_{3}$OCH$_{3}$ \\
568690.1 & CH$_{3}$OCH$_{3}$ \\
568690.3 & CH$_{3}$OCH$_{3}$ \\
568690.4 & CH$_{3}$OCH$_{3}$ \\
568741.6 & SO \\
568783.6 & CH$_{3}$OH \\
568999.3 & CH$_{3}$CN \\
569091.6 & $^{34}$SO$_{2}$ \\
569138 & U-line \\
569324.5 & CH$_{3}$OH \\
569486.9 & CH$_{3}$CN \\
569606.3 & CH$_{3}$CN \\
569704.0 & CH$_{3}$CN \\
569780.1 & CH$_{3}$CN \\
569834.5 & CH$_{3}$CN \\
569867.1 & CH$_{3}$CN \\
569878.0 & CH$_{3}$CN \\
570219.1 & CH$_{3}$OCH$_{3}$ \\
570221.9 & CH$_{3}$OCH$_{3}$ \\
570223.3 & CH$_{3}$OCH$_{3}$ \\
570223.3 & CH$_{3}$OCH$_{3}$ \\
570224.7 & CH$_{3}$OCH$_{3}$ \\
570224.7 & CH$_{3}$OCH$_{3}$ \\
570226.1 & CH$_{3}$OCH$_{3}$ \\
570227.5 & CH$_{3}$OCH$_{3}$ \\
570261.5 & CH$_{3}$OH \\
570264.0 & CH$_{3}$OH \\
\hline
\end{tabular}
\end{minipage}
\begin{minipage}[tc]{0.333\textwidth}
\begin{tabular}{ll}
\hline
\hline
Frequency & Molecule \\
{}[MHz] & or U/T-line \\
\hline
570303 & U-line \\
570335 & T-line \\
570619.0 & CH$_{3}$OH \\
570624.2 & $^{13}$CH$_{3}$OH \\
570790 & T-line \\
570814 & T-line \\
571119.7 & OCS \\
571151 & T-line \\
571217 & T-line \\
571477 & T-line \\
571532.6 & SO$_{2}$ \\
571553.3 & SO$_{2}$ \\
572112.8 & $^{15}$NH$_{3}$ \\
572498.2 & NH$_{3}$ \\
572596 & U-line \\
572676.2 & CH$_{3}$OH \\
572678 & U-line \\
572898.8 & CH$_{3}$OH \\
573471.1 & CH$_{3}$OH \\
573527.3 & $^{34}$SO$_{2}$ \\
573912.7 & CH$_{3}$OH \\
574090.9 & CH$_{3}$OCH$_{3}$ \\
574090.9 & CH$_{3}$OCH$_{3}$ \\
574093.3 & CH$_{3}$OCH$_{3}$ \\
574095.7 & CH$_{3}$OCH$_{3}$ \\
574140.0 & H$_{2}$CS \\
574184 & U-line \\
574587.8 & SO$_{2}$ \\
574797.9 & $^{34}$SO$_{2}$ \\
574807.3 & SO$_{2}$ \\
574868.5 & CH$_{3}$OH \\
575397 & T-line \\
575547.2 & CH$_{3}$OH \\
576042.1 & SO$_{2}$ \\
576267.9 & CO \\
576446 & U-line \\
576708.3 & H$_{2}$CO \\
576720.2 & NS \\
577160 & T-line \\
\hline
\end{tabular}
\vspace{14\baselineskip}
\end{minipage}
\end{table*}